\documentclass{article}
\usepackage{epsfig}
\usepackage{amsmath}
\usepackage{amsfonts}\tolerance=10000
\date{}

\newcommand{\ka}{\kappa}

\newcommand{\si}{\sigma}

\newcommand{\f}{\phi}

\newcommand{\F}{\Phi}

\newcommand{\ee}{\end{equation}}
\newcommand{\eea}{\end{eqnarray}}
\newcommand{\be}{\begin{equation}}
\newcommand{\bea}{\begin{eqnarray}}

\newcommand{\pa}{\partial}

\newcommand{\vep}{\varepsilon}

\newcommand{\re}[1]{(\ref{#1})}
\newcommand{\R}{{\rm I \hspace{-0.52ex} R}}

\begin{document} 

\title{Gravitating Yang--Mills fields in all dimensions } 
 
\author{
{\large Eugen Radu}$^{\dagger}$
and {\large D. H. Tchrakian}$^{\ddagger \star}$ \\ \\
$^{\dagger}${\small Institut f\"ur Physik, Universit\"at Oldenburg, Postfach 2503
D-26111 Oldenburg, Germany}
\\
$^{\ddagger}${\small  School of Theoretical Physics -- DIAS, 10 Burlington
Road, Dublin 4, Ireland}
\\
$^{\star}${\small Department of Computer Science,
National University of Ireland Maynooth,
Maynooth,
Ireland}
  }
 
\maketitle 
\begin{abstract}
A classification of gravitating Yang--Mills systems in all dimensions is presented. These systems are set up
so that they support finite energy solutions.
Both regular and black hole
solutions are considered, the former being the limit of the latter for vanishing event horizon radius.
  Special attention is paid to systems
necessarily involving higher order Yang--Mills curvature terms, along with the option of incorporating higher order terms
in the Riemann curvature.
The scope here is restricted
to Einstein systems, with or without cosmological constant, and the Yang--Mills(--Higgs) systems.
\end{abstract}
 
\section{Introduction}
By gravitating Yang--Mills (YM) fields we understand YM fields on curved backgrounds whose dynamics includes the
{\em backreaction} of the gravitational field on it. These are the particle-like and black hole solutions to
the Einstein-Yang--Mills (EYM) systems.

 Initially, EYM solutions~\cite{Bartnik:1988am,Volkov:1998cc} in
$3+1$ dimensional space with Minkowskian signature were primarily of interest because they presented configurations with
{\em non-Abelian hair}\footnote{These were preceded historically by gravitating Skyrmions~\cite{gravskyrme} which exhibited {\em Skyrmion hair}, but
the latter have not been studied as intensively since.}.
More recently however, gravitating non-Abelian solutions have found extensive application in
string inspired theories, {\it e.g.}, in various supergravity and $D-$brane models. These results point to the
physical relevance of classical
gravitating non-Abelian solutions. Moreover in this context, it is mainly such classical solutions in dimensions higher that
$3+1$ that find applications.
It is our aim here to review gravitating non-Abelian YM solutions in higher dimensions\footnote{It
 is concievable that gravitating Skyrmions may also be of relevance to field theories in higher dimensions, but to
date there has been
no work on them reported in the literature.}, together with their four dimensional counterparts.

Soon after the original particle-like (regular) solutions of \cite{Bartnik:1988am} were constructed, the corresponding
black hole solutions were found~\cite{Volkov:1989fi,Kuenzle:1990is,Bizon:1990sr}. Not long after the discovery of the EYM
solutions, both regular and black hole gravitating monopole solutions of the EYM--Higgs (EYMH) systems were
constructed in \cite{Breitenlohner:1991aa,Breitenlohner:1994di,Lee:1991vy}. EYM solutions in the presence of a
cosmological constant $\Lambda$ were constructed later, with negative $\Lambda$ in
\cite{Winstanley:1998sn,Bjoraker:1999yd,Bjoraker:2000qd}, and for $\Lambda>0$ in \cite{Volkov:1996qj}.
These results 
indicate  that the solutions of the Einstein equations coupled to non-Abelian
matter fields possess a much richer structure 
than in the better known U(1) case. They also show that our intuition based on solutions with linear field sources
may fail in more general situations.

In this review, our description of gravitating YM systems will in addition to the EYM fields also include the Higgs
fields. The study of EYM-Higgs (EYMH) systems enables a more extensive description of physical phenomena as a result of
the symmetry breaking
mechanism which (gauged) Higgs models describe. This is a result of the special nature of Higgs fields  here  as
dimensional descendents of gauge fields, as will be explained below. 
Therefore we shall restrict our considerations to such Higgs
multiplets that afford topological stability to the solutions.

EYM solutions in (higher) $d>4$ dimensions were considered relatively recently. 
Here two main possibilities have been included so far in the literature, which are distinguished by 
the different asymptotic structure of the spacetime.
In the first case, of main interest here, the metric is asymptotically Minkowski (or (anti-)de Sitter, if a cosmological
constant is considered in the action). 
Such solutions provide the natural counterparts of the $d=3+1$ non-Abelian configurations mentioned above. In the second case,
a number of $p$ codimensions are included. Such configurations are important if one posits 
the existence of extra-dimensions in the universe, which are likely to be compact and described by a Kaluza-Klein (KK) theory. 
All known KK non-Abelian solutions have no dependence on the extra-dimensions, $i.e.$ these are frozen,  as in the case of the $z-$coordinate
of the Abrikosov-Nielsen-Olesen vortex.

A number of results in the literature show that, in the absence of codimensions,
the mass of gravitating non-Abelian solutions which asymptote to a Minkowski spacetime, is infinite \cite{Volkov:2001tb}
(the same results holds for asymptotically (anti-)de Sitter solutions \cite{Okuyama:2002mh}).
This is not surprising because the usual EYM system in $d\ge 5$ dimensions does not have the
requisite scaling properties for there to exist finite energy solutions.
 
However, the scaling properties of the usual EYM system can be altered by the addition of higher order terms of the
Yang--Mills curvatures. Such terms can occur in the
low energy effective action of string theory \cite{GSW}, \cite{Pol}.
The hierarchies of both YM and of gravitational ($i.e.$, Einstein) systems will be defined below.
Employing suitably defined EYM systems featuring higher order YM curvature~\footnote{Higher order Riemann curvature
terms cannot be employed for this purpose for when they are of sufficiently higher order to satisfy the scaling
requirement, they invariably vanish or are $total\ divergence$ terms.} terms such that
the equations of motion remain second order, 
finite mass/energy static spherically symmetric non-Abelian solutions in $d>4$ dimensions were constructed. 
The existence of this type of configurations is a nonperturbative effect, since they cannot be predicted in
a perturbative approach around the   solutions of the usual  EYM system.
 
With zero cosmological canstant, $d>4$ regular solutions 
of the extended EYM system were found and analysed in \cite{Brihaye:2002hr,Breitenlohner:2005hx}, and black hole solutions in 
\cite{Brihaye:2002jg}, \cite{Radu:2005mj}.  Both regular and black hole with negative cosmological canstant $\Lambda$ were
presented in \cite{Radu:2005mj}. With positive $\Lambda$ likewise, such EYM solutions were given
in \cite{Brihaye:2006xc}. Thus there is a comprehensive sample of finite mass, static, spherically symmetric EYM
field configurations in $d=D+1$ dimensional spacetimes. These are the higher dimensional counterparts of the EYM
solutions in four dimensions, and in general their non-Abelian matter content is composed of several
YM terms with various (appropriate) scaling properties. Such models proliferate with increasing dimension and
exhibit additional features, absent in the usual $3+1$ dimensional EYM case\footnote{The exception is a very particular
hierarchy of gravitating YM models in $d=4p$ dimensions, consisting of a single higher order YM curvature term which scales
appropriately. In this case~\cite{Radu:2006mb} all the qualitative features of the EYM solutions in $d=4$ are preserved.
}.

Concerning EYMH solutions describing gravitating monopoles in higher dimensions, recently a very
particular hierarchy of YMH models in $d=4p$ dimensions has been studied where regular and black hole solutions are
constructed~\cite{Breitenlohner:2009zi}. The reason for restricting to a particular
family of models is the ubiquity of Higgs models in dimensions higher than $3+1$, as will be described below.

All the above noted EYM and EYMH solutions are static and spherically symmetric. But in $3+1$ dimensions there are axially
symmetric solutions~\cite{Kleihaus:1996vi,Kleihaus:1997ic,Hartmann:2000gx}
with very interesting properties, so it is natural to seek their higher dimensional counterparts. 
To the best of our knowledge, the only result in this direction is that in \cite{Radu:2007jb},
for the simplest EYM system in $d=4+1$ dimensions. The symmetry imposed in that case was bi-azimuthal symmetry in the
$4$ spacelike dimensions.

Concerning the case of non-Abelian solutions with codimensions, the situation is less explored, the case
$d=5$ being the only one discussed in a systematic way in the literature.
Both non-Abelian vortices and black strings with non-Abelian hair have been studied in the literature, starting with the 
pioneering work of \cite{Volkov:2001tb}. 

Finally, we note that all the work referred to above pertains to gravitating YM in $D+1$ dimensional spacetime with
Minkowskian signature. 
This is because to date all of the work on Euclidean EYM in all
dimensions is effectively restricted to the study of YM fields on fixed (gravitational) backgrounds. This,
even though the earliest EYM solutions were constructed in $4$ Euclidean dimensions
(see \cite{Charap:1977re,Charap:1977ww} and \cite{BoutalebJoutei:1979va,Chakrabarti:1987kz}). As such, these results are
not central to the present review which is mainly concerned with fully backreacting matter-gravity solutions.

The outline of this paper is as follows.
In the next section, {\bf 2}, we introduce the hierarchies of gravitational and Yang--Mills (and Higgs) systems,
in all dimensions. Then in the following section, {\bf 3}, we review known results in gravitating YM and
YMH solutions in higher dimensions, including a brief description of EYM solutions in Euclidean space.
There we pay special attention to the existence and the generic properties of the non-Abelian solutions
and not to their physical applications. In section {\bf 4} we give a summary and outlook.

\section{Einstein and Yang--Mills hierarchies}
The gravitating solutions of interest are those with finite energy in the Minkowskian regime,
and with finite action in the Euclidean case.
 They would describe spherically symmetric black hole solutions of horizon radius $r_H$, which include the regular
particle like configuration in the limit of $r_H=0$. As noted in section {\bf 1}, the non-Abelian matter sector will consist
both of YM, and YM-Higgs (YMH) fields. These  systems are defined on Euclidean spaces since in the Minkowskian case the
solutions in question are static so that the fields depend on the spacelike coordinates only.

Even before gravitating the YMH field, the existence of finite energy/action solutions is contingent on the requisite
scaling requirement being satisfied. In the case of flat (Euclidean) space, this amounts to satisfying the familiar
virial relation, as long as the (static) Hamiltonian/Lagrangian is positive definite. In the case of a
gravitating system the property of positive definiteness is absent, since all Einstein systems are defined such that they
are not bounded from below. One can procede nonetheless heuristically seeking to satisfy the virial relation.
A systematic analysis of this for higher dimensional EYM systems was presented in \cite{Radu:2005mj}, and we will
return to this question in section {\bf 3} when the static systems are gravitated and will elaborate on the scaling
arguments. For now, we restrict our attention to the YMH systems on a flat backround. It is the virial constraint that
necessitates the introduction of higher order curvature terms in the (static) Hamiltonian/Lagrangian. These are the
members of the Einstein and YM hierarchies to be defined below.

There is a marked difference in the status of the Einstein hierarchy and the YM hierarchy in the present context.
The higher order YM terms are necessary for rendering the scaling properties of the EYM system in question appropriate
for supporting finite energy solutions, while the correspoding terms in the gravitational hierarchy do not play this role.
All the higher order  gravitational terms ({\it e.g.}, Gauss--Bonnet) with the required scaling property either vanish
or are $total\ divergence$.

Anticipating the definition of the Einstein hierarchy in the next subsection, we note that
the $p-$Einstein system $e\,R_{(p)}$ as defined by the relations \re{gp}, or \re{ricci-s} bellow, 
scales as $L^{-2p}$. (Note that the usual
Einstein-Hilbert term,( or $1-$Einstein system in the terminology of this work), 
scales as $L^{-2}$ and the usual YM, or $1-$YM system, scales as $L^{-4}$.)
According to the heuristic scaling argument, adding the $p-$Einstein term to the usual EYM system would result in the
correct scaling if $d<2p+1$. In the limiting case when $d=2p$, $e\,R_{(p)}$ is a $total\ divergence$ and beyond that
it vansihes, as will be seen from the definitions below.

On the other, hand higher order members of the gravitational hierarchy play
a quantitatively interesting role in highlighting certain qualitative features of the solutions, that
repeat in dimensions {\it modulo} $4p$. They are also of intrinsic interest in some considerations for the case of
Euclidean signature. For those reasons, they are included in this review.

In the next subsection, {\bf 2.1}, we present the definition of gravitational systems in all dimensions, which in this
review we refer to as the hierarchy of Einstein systems. In general, higher dimensional gravitational systems are
composed of the superposition of individual members of the Einstein hierarchy.
In subsection {\bf 2.2} we define the Yang--Mills (YM)
systems in all even (spacelike) dimensions, referred to as the hierarchy of YM systems here. YM systems in all dimensions,
both odd and even, can then be constructed from the superposition of individual members of the YM hierarchy.
Then in subsection {\bf 2.3}
we introduce that subclass of (gauged) Higgs models that are relevant to the presentation here, namely those YM--Higgs
(YMH) models that have been gravitated to date, along with an example of the next most natural candidate. Finally
in subsection {\bf 2.4} we state the expressions for the reduced Lagrangians of the systems introduced, subject to the
appropriate symmetries. This will be mainly the case of most interest involving the static fields which are spherically
symmetric in the spacelike dimensions.

\subsection{Gravitational systems in all dimensions: Einstein hierarchy}

The gravitational systems which we shall refer to as the Einstein hierarchy in $d=2p+k$ spacetime dimensions
are defined as the product of the determinant of the {\it Vielbeine} $e_{\nu}^{n}$, $e=\mbox{det}\,e_{\mu}^{m}$,
with Ricci scalars $R_{(p)}$, defined by
\be
\label{gp}
{\cal L}_{p-{\rm E}}=e\,R_{(p)}=\vep^{\mu_1\mu_2\dots\mu_{2p}\nu_1\nu_2\dots\nu_{k}}
e_{\nu_1}^{n_1}e_{\nu_2}^{n_{2q}}\dots e_{\nu_{k}}^{n_{k}}
~\vep_{m_1m_2...m_{2p}n_1n_2\dots n_{k}}\,
R_{\mu_1\mu_2\dots\mu_{2p}}^{m_1m_2\dots m_{2p}}\,,
\ee
$R_{\mu_1\mu_2\dots\mu_{2p}}^{m_1m_2\dots m_{2p}}=R(2p)$ being the $p$-fold totally
antisymmetrised product of the Riemann curvature, in component notation
\[
R(2)=R_{\mu_1\mu_2}^{m_1m_2}=\pa_{[\mu_1}\,\omega_{\mu_2]}^{m_1m_2}+\omega_{[\mu_1}^{m_1n}\,\omega_{\mu_2]}^{nm_2}\,,
\]
$\omega_{\mu}^{mn}$ being the Levi-Civita spin connection. It is
clear from the definition \re{gp} that the spacetime dimensionality $d=2p+k$ sets an upper limit on the highest
order nontrivial member of this hierarchy, the term $R_{(p)(k=0)}$ being the (total divergence)
Euler-Hirzebruch density. We shall refer to $e\,R_{(p)}$  in \re{gp} as $p-$Einstein systems 
($i.e.$ $p=1$ is the usual Einstein-Hilbert Lagrangean, $p=2$ is the Gauss-Bonnet term etc.).

Subjecting \re{gp} to the variational principle with respect to the arbitrary variation of
the {\it Vielbeine} one arrives at what we refer to as the $p$-Einstein equation
\be
\label{peinstein}
G_{(p)}{}_{\mu}^m=R_{(p)}{}_{\mu}^m\
-\ \frac{1}{2p}\ R_{(p)}\ e_{\mu}^m\,,
\ee
in terms of the $p$-Einstein tensor $G_{(p)}{}_{\mu}^m$, with $R_{(p)}$ and $R_{(p)}{}_{\mu}^{m}$ being the $p$-th order
Ricci scalar and the $p$-th order Ricci tensor defined respectively by
\bea
R_{(p)}&=&R_{\mu_1\,\mu_2\dots\mu_{2p}}^{m_1\,m_2\dots m_{2p}}\;
e^{\mu_1}_{m_1}\,e^{\mu_1}_{m_2}\dots e^{\mu_{2p}}_{m_{2p}}\label{ricci-s}\\
R_{(p)}{}_{\mu}^{m}&=&R_{\mu\mu_2\mu_3\dots\mu_{2p}}^{m\,m_2\,m_3\dots m_{2p}}\;
e^{\mu_2}_{m_2}\,e^{\mu_3}_{m_3}\dots\,e^{\mu_{2p}}_{m_{2p}}\;\;.\label{ricci-t}
\eea

The $p-$Einstein hierarchy of gravitational systems is defined by $e$ times the $p-$Ricci scalar \re{ricci-s}, and the most
general Einstein system is given by the maximal number of nonvanishing superpositions of all $p-$Ricci scalars.

An interesting property of the $p-$Riemann and $q-$Riemann curvatures in even dimensions is the double-selfduality
constraint
\be
\label{dsd}
R_{\mu_1\mu_2....\mu_{2p}}^{m_1m_2..m_{2p}}=\pm\,\ka^{2(p-q)}
\frac{e}{[(2q)!]^2}\,
\vep_{\mu_1\mu_2...\mu_{2p}\nu_1\nu_2...\nu_{2q}}
R^{\nu_1\nu_2....\nu_{2q}}_{n_1n_2..n_{2q}}\,
~\vep^{m_1m_2...m_{2p}n_1n_2...n_{2q}}\,,
\ee
where the Hodge dual of the $2q-$form curvature is equated to the $2p-$form curvature, with the dimensionful
constant $\ka$ compensating for the difference in the respective dimensions. \re{dsd} can be stated for both
Euclidean and Minkowskian signatures, with the $\pm$ sign respectively.For Euclidean signature, contracting the constraint \re{dsd} with the appropriate number of {\it vielbeine} one arrives at the vacuum Einstein equations for the $(p,q)-$Einstein system
\be
\label{pqeinstein}
{\cal L}_{(p,q)-{\rm E}}=e\,(R_{(p)}+\ka^{2{(p-q)}}R_{(q)}+\Lambda)\,,
\ee
where $R_{(p)}$ and $R_{(q)}$ in \re{pqeinstein} are defined by \re{ricci-s} and $\Lambda$ is a cosmological constant whose value is
related to the constant $\ka$, as shown in \cite{O'Brien:1988rs}. This is not valid in the case of Minkowskian signature
(see Appendix of \cite{Radu:2007az} for details).

\subsection{The Yang--Mills hierarchy}

Since we seek finite energy/action solutions of the gravitating YM systems, the relevant members of the YM hierarchy
to be defined, are those which support finite action solutions in the spacelike (Euclidean) subspace $D$ of the spacetime
$d=D+1$. The flat space solutions in question will have topological stability when $D$ is even. To avail of topological
stability for these finite energy/action solutions in {\bf odd} dimensions $D$, one has to consider the Higgs
models derived from the dimensional descent of the YM systems in {\bf even} $D+N$ with compact {\bf odd} codimension
$N$. The YM hierarchy is presented in the section {\bf 2.2.1} and the ensuing YMH models in section {\bf 2.2.2}, below.

Using the notation $F(2)=F_{\mu\nu}$ for the $2$-form Yang--Mills (YM)
curvature, the $2p$-form YM tensor
\be
\label{2p}
F(2p)=F(2)\wedge F(2)\wedge...\wedge F(2)\ ,\quad p-{\rm times}
\ee
is a $p$ fold totally antisymmetrised product of the $2$-form curvature.

The $p-$YM system of the YM hierarchy is defined, on $\R^{4p}$, by the Lagrangian density
\be
\label{pYM}
{\cal L}_{p-{\rm YM}}=\mbox{Tr}F(2p)^2\,.
\ee

In $2n$ dimensions, partitioning $n$ as $n=p+q$, the Hodge dual of the
$2q$-form field $F(2q)$, namely $(^{\star}F(2q))(2p)$, is a $2p$-form.

Starting from the inequality
\be
\label{ineq}
\mbox{Tr}[F(2p)-\ka\ \ ^{\star}F(2q)]^2\ge 0\ ,
\ee
it follows that
\be
\label{top-lb1}
\mbox{Tr}[F(2p)^2+\ka^2\ F(2q)^2]\ge 2\ka\ {\cal C}_n\ ,
\ee
where ${\cal C}_n$ is the $n$-th Chern-Pontryagin density. In \re{ineq}
and \re{top-lb1}, the constant $\ka$ has the dimension of length to the
power of $(p-q)$.

The element of the YM systems labeled by $(p,q)$ in (even) $2(p+q)$ dimensions are defined
by Lagrangians defined by the densities on the {\it left hand side} of
\re{top-lb1}. When in particular $p=q$, then these systems are
conformally invariant and we refer to them as the $p-$YM members of the YM hierarchy.

The inequality \re{top-lb1} presents a topological lower bound which
guarantees that finite action solutions to the Euler--Lagrange equations
exist. Of particular interest are solutions to first order self-duality
equations which solve the second order Euler--Lagrange equations, when
\re{top-lb1} can be saturated.

For ${\bf M}^{2n}={\R}^{2n}$, the self--duality equations support
nontrivial solutions only if $q=p$,
\be
\label{sd-4p}
F(2p)\ =\ ^{\star}F(2p)\ .
\ee
For $p=1$, i.e. in four Euclidean dimensions, \re{sd-4p} is the usual
YM selfduality equation supporting instanton solutions. Of these, the
spherically symmetric~\cite{Belavin:1975fg,Tchrakian:1984gq} and axially
symmetric~\cite{Witten:1976ck,Chakrabarti:1985qj,Spruck:1997eb} instantons on $\R^{4p}$
are the known. For $p\ge 2$, i.e. in dimensions eight and higher, only
sphericaly symmetric~\cite{Tchrakian:1984gq} and axially symmetric~\cite{Chakrabarti:1985qj,Spruck:1997eb} solutions
can be constructed, because in these dimensions \re{sd-4p} are
overdetermined~\cite{Tchrakian:1990gc}.

In the $r\gg 1$ region, all these 'instanton' fields on ${\bf R}^{2n}$,
whether self--dual or not, asymptotically behave as pure--gauge
\[
A\rightarrow gdg^{-1}
\]
For ${\bf M}^{2n}=G/H$, namely on compact coset spaces, the self--duality
equations support nontrivial solutions for all $p$ and $q$,
\be
\label{sd-g/h}
F(2p)\ =\ \ka\ ^{\star}F(2q)\ 
\ee
where the constant $\ka$ is some power of the 'radius' of the (compact)
space. The simplest examples are ${\bf M}^{2n}=S^{2n}$, the
$2n$-spheres~\cite{O'Se:1987fx}, and ${\bf M}^{2n}={\bf CP}^n$, the complex
projective spaces~\cite{Ma:1990ja}.

The above definitions of the YM systems can be formally extended to all dimensions, including all odd dimensions.
The only difference this makes is that all topological lower bounds enabling the construction of instantons are then
lost, but this is immaterial from the viewpoint in the present review.

\subsection{Higgs models on ${\bf R}^D$}

Higgs fields have the same dimensions as gauge connections and appear
as the extra components of the latter under dimensional reduction, when
the extra dimension is a compact symmetric space. Dimensional reduction of gauge fields over a compact
codimension is implemented by the imposition of the symmetry of the compact coset space on the
coordinates of the codimensions. In this respect, the calculus of dimensional reduction does not
differ from that of imposition of symmetries generally, which is the relevant formalism used in this,
{\bf 2.3}, and the next subsection {\bf 2.3}.

The calculus of imposition of symmetry on gauge fields that has been used in the works being reviewed here
is that of Schwarz~\cite{Schwarz:1977ix,Romanov:1977rr,Schwarz:1981mb}. This formalism was adapted to
the dimensional reduction over arbitrary codimensions in \cite{Ma:1986pu,Ma:1988um,O'Brien:1988rs}.

In general one can employ a linear combination of inequalities \re{top-lb1}, for all
$p\le D/4$ and $q\le D/4$. Restricting, for simplicity
to the $4p$ dimensional conformal invariant systems in
\re{top-lb1}, i.e. to $p=q=D/4$, the descent over the compact space
$K^{4p-d}$ is described
by
\be
\label{descent}
\int_{{\R}^D\times K^{4p-D}}{\cal F}(2p)^2\ \ge
\int_{{\R}^D\times K^{4p-D}}{\cal C}_{2p}\ ,
\ee
where ${\cal F}(2p)$ is the $2p-$form curvature of the $1-$form connection ${\cal A}$
on the higher dimensional space ${\bf R}^D\times K^{4p-D}$.
Imposing the symmetry appropriate to $K^{4p-D}$ on the gauge fields
results in the breaking of the original gauge group to, say, the
residual gauge group $g$ for the fields on ${\bf R}^D$. Performing
then the integration over the compact space $K^{4p-D}$ leads to the Lagrangian
${\cal L}[A,\f]$, of the residual Higgs model on ${\bf R}^D$.
$A$ here is the connection taking values in the algebra of $g$
and $\phi$ is the Higgs multiplet whose structure under $g$ depends
on the detailed choice of $K^{4p-D}$, implying the following gauge
transformations
\[
{A}\rightarrow g{A} g^{-1}+gd\ g^{-1}
\]
and depending on the choice of $K^{4p-D}$,
\[
\phi\rightarrow g\phi\ g^{-1}\quad ,\quad{\rm or}\quad ,\quad
\phi\rightarrow g\phi\quad ,\quad{\rm etc.}
\]

The inequality \re{top-lb1} leads, after this dimensional descent, to
\bea
\int_{{\R}^D}{\cal L}[{A},\f]&\ge&\int_{{\R}^D}\nabla\cdot
{\bf \Omega}[{A},\phi]=\int_{\bf\Sigma^{D-1}}{\bf\Omega}[{A},\phi]\ ,\label{top-lb2b}
\eea
where ${\cal L}[{A},\phi]=
{\cal L}[{F},{D}\phi,\vert\phi\vert^2,\eta^2]$ is the residual
Lagrangian in terms of the residual gauge connection ${\cal A}$ and its
curvature $F$, the Higgs fields $\phi$ and its covariant derivative
$D\phi$ and the inverse of the compactification 'radius' $\eta$.
The latter is simply the VEV of the Higgs field, seen clearly from the
typical form of the components of the curvature $F$ on the extra (compact)
space $K^{4p-D}$
\be
\label{VEV}
F\vert_{K^{4p-D}}\ \sim\ (\eta^2-\vert\phi\vert^2)
\otimes\Sigma\quad \Rightarrow
\quad\lim_{r\to\infty}\vert\phi\vert^2=\eta^2
\ee
where $\Sigma$ are, symbolically, spin-matrices/Clebsch-Gordan coefficients.

It should be noted at this stage that subjecting the selfduality equations \re{sd-4p} to this dimensional descent
results in Bogomol'nyi equations on $\R^D$, which for $p\ge 2$ in ${\R}^D\times K^{4p-D}$ ($cf.$ \re{descent})
turn out to be overdetermined~\cite{Tchrakian:1990gc} with few exceptions.

There arise a plethora of Higgs models, depending on the mode of dimensional descent, namely on the particular
choice of the compact codimension $K^{4p-d}$. We will not dwell on various modes of descent and the detailed properties
of the descended YMH models here, and will limit our attention to those models that have been gravitated to date.

Perhaps the most interesting, or useful, family of YMH models on $\R^D$ arrived at $via$ this descent mechanism are those
in which the residual gauge group is $SO(D)$, and the Higgs field multiplet is an isovector of
$SO(D)$~\footnote{Emlpoying Dirac matrix representations for the algebra of $SO(D)$ in terms of
$\Gamma_{ij}$, $i,j=1,2,\dots,$, the Higgs field takes its values in the matrix basis $\Gamma_{i,D+1}$. with
$(\Gamma_{ij},\Gamma_{i,D+1})$ representing the algebra of $SO(d)=SO(D+1)$}. It turns out
that only when $D=2$ and when $D=4p-1$ in the descent over ${\R}^D\times K^{4p-D}$, the resulting Bogomol'nyi equations
are {\bf not} overdetermined~\cite{Tchrakian:1990gc}. The $D=2$ case is uninteresting from our present perspective since
the Abelian Higgs systems in that case live in $2+1$
spacetime dimensions and gravitating them is unproductive. This leaves the family of $SO(D)$ Higgs models
that live in $D=4p-1$ space, or $d=4p$ spacetime dimensions, for which the flat space Bogomol'nyi equations can be
saturated. These are the only Higgs models to date, that are gravitated~\cite{Breitenlohner:2009zi}. (The flat
space solutions of this system was studied in \cite{Radu:2005rf}, which are direct generalisations of the usual BPS
monopoles with $p=1$.)

The YM field in the YMH models on $R^D$ discussed thus far, is purely magnetic supporting a 'magnetic' monopole. But 
when it comes to YMH models, as stated earlier, the presence of the Higgs field enables the support of a dyon
in $d=D+1$ dimensional spacetime. In the usual~\cite{Julia:1975ff} sense as the dyon in $3+1$ dimensions, the Higgs
field partners the newly introduced 'electric' YM potential $A_0$. Thus we can describe $SO(d)$ dyons~\footnote{While
both the Higgs field and $A_0$ take their values in the Dirac matrix basis $\Gamma_{i,D+1}$, in the
in the $3+1$ dimensional case the 'enlarged' algebra $SO(4)$ splits in the two $SU(2)$ subalgebras, whence
the dyon and the monopole are both described by $SU(2)$ matrices. In all higher dimensions, this is not the case and the
full algebra employed is that is $SO(D+1)$, where $d=D+1$ is the dimension of the spacetime.} in $d-$dimensional
spacetime.

When the dimension of the spacetime $d$ is even, then the chiral representations of the algebra of $SO(d)$ can be
employed, namely replacing the Dirac representation matrices $\Gamma_{\mu\nu}=(\Gamma_{ij},\Gamma_{i,d})$
with $\Sigma_{\mu\nu}^{(\pm)}=(\Sigma_{ij}^{(\pm)},\Sigma_{i,d}^{(\pm)})$, the precise definition of these matrices to be
given explicitly in the next subsection.

When by contrast the dimension of the spacetime $d$ is odd, then the Dirac represenation matrices are the
appropriate ones to be used, $e.g.$, in $d=4+1$ spacetime~\cite{O'Brien:1988xr} when $D=4$. (No such higher dimensional
monopoles are gravitated to date.)

The family of YMH in $d=4p$ dimensional spacetime that are gravitated result from the simplest mode of descent over
${\R}^D\times K^{1}$, with $D=4p-1$, $i.e.$, over one codimension. The Lagrangian densities can be expressed for
arbitrary $p$ in flat space as~\cite{Radu:2005rf}
\bea
{\cal L}_{p-{\rm YMH}}&=&\mbox{Tr}\,\left[F(2p)^2+2p\,
\left(F(2p-2)\wedge D\Phi\right)^2\right]\nonumber\\
&=&\mbox{Tr}\,\left[(F_{m_1m_2...m_{2p}})^2+2p\,
\left(F_{[m_1m_2...m_{2p-2}}\,D_{i_{2p-1}]}\Phi\right)^2\right]\label{4pres}\,,
\eea
in an obvious notation.

\subsection{Static spherically symmetric fields}
Since almost all the work reviewed in this article involves static spherically symmetric only, we will subject the
above introduced systems to this symmetry only.

The usual metric Ansatz with spherical symmetry in $d-1$ dimensional subspace is
\be
\label{metric}
ds^2 = \mp\,N(r)\sigma^2(r) d\tau^2\ +\ N(r)^{-1} dr^2\
+\ r^2 d \Omega_{(d-2)}^2 \: ,
\ee
the $\mp$ sign pertaining to Lorentzian and Euclidean signatures, and with $d \Omega_{(d-2)}^2$ being the
metric on $S^{d-2}$.

Subject to the Ansatz \re{metric}, the reduced one dimensional Lagrangian of the $p-$Einstein system \re{gp}
(or \re{ricci-s}), in $d-$dimensional spacetime is calculated. After neglecting the appropriate surface terms,
this can be expressed compactly as
\be
\label{redlaggrav}
L^{(p,d)}_{(\rm{grav})}=
\frac{\ka_p}{2^{2p-1}}\,\frac{(d-2)!}{(d-2p-1)!}\ \si\,
\frac{d}{dr}\left[r^{d-2p-1}(1-N)^p\right]\,.
\ee
We next state the static spherically symmetric Ansatz for the $p-$YM system \re{pYM} and the $p-$YMH system,
\begin{equation}
\label{YMHsphodd}
\F=\eta\,h(r)\,\hat x_j\,\Gamma_{j,d}\ ,\quad
A_0=u(r)\,\hat x_j\,\Gamma_{j,d}\ ,\quad
A_i=\left(\frac{1-w(r)}{r}\right)\,\Gamma_{ij}\hat x_j\ .
\end{equation}
for odd dimensional spacetime $d$, and, for even $d$
\begin{equation}
\label{YMHspheven}
\F=\eta\,h(r)\,\hat x_j\,\Sigma_{j,d}^{(\pm)}\ ,\quad
A_0=u(r)\,\hat x_j\,\Sigma_{j,d}^{(\pm)}\ ,\quad
A_i=\left(\frac{1-w(r)}{r}\right)\,\Sigma_{ij}^{(\pm)}\hat x_j\ ,
\end{equation}
where
\[
\Sigma_{ij}^{(\pm)}=
-\frac{1}{4}\left(\frac{1\pm\Gamma_{d+1}}{2}\right)
[\Gamma_i ,\Gamma_j]\ ,
\]
are the chiral representations of $SO(d)$. The dimensionful constant $\eta$ in \re{YMHsphodd}-\re{YMHspheven} is the
Higgs VEV.

It should be stated here that the choice of gauge group here is made with the purpose of enabling the construction of
nontrivial finite energy solutions, and that this is the minimal size of gauge group in each case. Larger gauge groups, with
the appropriate representations containing these, can be employed in case of necessity, {\it e.g.}, when a Chern--Simons
term is to be introduced in the Lagarangian.

Imposing this symmetry, $i.e.$, substituting \re{YMHsphodd} or \re{YMHspheven} into the appropriate Lagrange density,
results in exactly the same one dimensional reduced Lagrangians in both cases (except for an overall factor of $2$).
This is because the algebraic manipulations involved in both cases are identical. The situation changes if Fermions
are introduced, but we do not do that here.

The resulting reduced one dimensional Lagrangian of the $p-$YM system \re{pYM}, augmented by the Higgs kinetic term in
\re{4pres}, in $d-$spacetime dimensions is
\bea
L_{\rm YMH}^{(p,d)}&=&\frac{\tau_p\,r^{d-2}}{2\cdot (2p)!}\Bigg\{
\frac{(d-2)!}{(d-[2p+1])!}\,\si
\left(\frac{1-w^2}{r^2}\right)^{2(p-1)}
\left[(2p)N\left(\frac{w'}{r}\right)^2+
(d-[2p+1])\,\left(\frac{1-w^2}{r^2}\right)^2\right]\nonumber\\
&\mp&\frac{(d-2)!}{(d-2p)!}\frac{(2p-1)}{\si}\left[([(1-w^2)^{p-1}u]')^2+(d-2p)\frac{(2p-1)}{N}[(1-w^2)^{p-1}u]^2\,\
\left(\frac{w}{r}\right)^2\right]\nonumber\\
&\mp&\frac{(d-2)!}{(d-2p)!}\frac{(2p-1)}{\si}\ \eta^2\left[([(1-w^2)^{p-1}h]')^2
+(d-2p)\frac{(2p-1)}{N}[(1-w^2)^{p-1}h]^2\,\
\left(\frac{w}{r}\right)^2\right]\Bigg\}\label{ymhp}
\eea
This is the most general matter part consisting of the YMH system for this {\it particular} family of Higgs models. 

Before proceeding to describe the various types of gravitating YM and YMH solutions in the next section, let us make
some remarks concerning the particular choices of the various models employed.

\begin{itemize}
\item
To
recover the formula used for the $p-$YM system \re{pYM}, one simply replaces $h=0$ in the third line of \re{ymhp},
thus eliminating the Higgs field.
\item
To recover the matter Lagrangian used for the gravitating monopoles in $d=4p$ dimensions presented in
\cite{Breitenlohner:2009zi},  one replaces $u\to 0$ in the second line of \re{ymhp}, and, sets $d=4p$ since the
gravitating monopoles there are constructed only in those spacetime dimensions. Of course, it would be possible also to
gravitate dyons in higher dimensions just as in $d=4$ (see for example \cite{Ibadov:2007qt} and references therein), but
this has not been done to date.

The choice of model for the monopoles gravitated in \cite{Breitenlohner:2009zi} was made firstly such that the YM term and
the Higgs kinetic term have the same dimensions. This is precisely with the criterion that there should not be a
mismatch of dimensions giving rise to a $conical\ fixed\ point$ sigularity, so as not to cloud an otherwise more
complicated system. This family of models is dimensionally descended from the $p-$YM system and hence the
Bogomol'nyi equations can be saturated, which is unimportant since the backreaction of gravity prevents this
saturation anyway. Secondly, we opted for gravitating  the $p-$Higgs models with $p-$gravity for the more or less
$aesthetic$ reasons of that choice in the EYM case of \cite{Radu:2006mb}.

\item
In the $d\geq 5$ EYM case, various matter systems consisting of the superpositions of $p-$YM systems \re{pYM} are gravitated
with the $1-$Einstein  gravity (usual Einstein-Hilbert Lagrangian). This immediately introduces a mismatch between the dimensions of the
constituent terms in the Lagrangian. It is found that one result of this is the absence of the radial excitations
(higher node solutions) observed in the
$p=1,~d=4$ Bartnik-McKinnon~\cite{Bartnik:1988am} case. Another result is that in
addition to the Reissner-Nordstr\"om fixed point, there arises a new singularity~\cite{Brihaye:2002jg} in $d=4+1$
dimensions. This singularity was found~\cite{Breitenlohner:2005hx} to repeat in $d=4p+1$ dimensions, $modulo$
$4p$. The fixed point corresponding to it was called a $conical\ fixed\ point$.

\item
The electric potential necessarily  vanishes for asymptotically flat 
finite energy solutions of the EYM system, $i.e.$ $u(r)=0$ if $h(r)=0$.
The proof here is similar to that found in \cite{Galtsov:1989ip}, \cite{Bizon:1992pi} for $d=3+1$ dimensions. 
One starts with the equation for the electric potential $u(r)$, which, for a generic model with a number of $P$ terms in
the YM ierarchy 
can be rewritten as
\begin{eqnarray}
\label{equ}
\sum_{p=1}^P \tau_p \frac{1}{2}\left(\frac{r^{d-2}}{\sigma} (W^2u^2)'\right)'
=\sum_{p=1}^P  \tau_p \frac{r^{d-2}}{\sigma} 
\left(((Wu)')^2+ \frac{2(d-2p)(2p-1)}{N}W^2\frac{u^2 w^2}{r^2}
\right),
\end{eqnarray}
(here, to simplify the relations, we denote $W=(1-w^2)^{p-1}$).
One can easily see that the r.h.s. of (\ref{equ}) is a stricly positive quantity.
Thus the integral of the  l.h.s. 
should also be positive,
\begin{eqnarray}
\sum_{p=1}^P \tau_p \frac{1}{2}\left(\frac{r^{d-2}}{\sigma} (W^2u^2)'\right)\bigg|_{r_0}^{\infty}
>0~,
\end{eqnarray}
(where $r_0=0,r_H$ for particle-like 
and black hole solutions, respectively).
However, the regularity of the solutions together with finite energy requirements impose that, in the above relation, both the contributions
at $r=r_0$ and at infinity vanish. As a result, $u(r)$ should vanish for any reasonable solution. 
The same proof generalises for anti-de Sitter solutions, the only exception  being the systems
featuring exclusively the $p$-th terms of the hierarchy, in $d=4p$ dimensions.

\item
Departing from these generic models, there is a family of models for which the mismatch of the dimensionality of the
constituent terms is removed. Like in the usual EYM system consisting of the $1-$Einstein and $1-$YM systems in $d=3+1$,
this family of models~\cite{Radu:2006mb} consists exclusively of the $p-$Einstein and $p-$YM systems in $d=4p$.
The result is that all qualitative features of the EYM solutions of \cite{Bartnik:1988am} are preserved. Indeed, if
$p-$Einstein is replaced by $q-$Einstein, $q\neq p$, the salient features persist but are quantitatively somewhat
deformed.

\end{itemize}

\section{Non-Abelian solutions in $d-$dimensions}

\subsection{Solutions with Lorentzian signature}

\subsubsection{ Einstein--Yang-Mills solutions in four dimensions}
The closed form solutions in Chakrabarti {\it et al}~\cite{BoutalebJoutei:1979va}, \cite{Chakrabarti:1987kz} 
are probably the first examples of black holes with non-Abelian hair, albeit with no backreaction 
between gravity and the non-Abelian matter. Fully selfgravitating EYM solutions were constructed somewhat after those on fixed backgrounds,
by Bartnik and McKinnon~\cite{Bartnik:1988am}. These were regular particle-like
solutions, and were soon followed by their black hole 
counterparts in \cite{Volkov:1989fi,Kuenzle:1990is,Bizon:1990sr}.

Subsequently,  a large literature has developed on this subject, extending to systems with a cosmological constant, 
and separately, to systems whose Lagrangian contains also a Higgs field, supporting gravitating monopoles. 
Extensions of the EYM system to include other fields which enter various stringy models 
have been considered as well, in particular for a Gauss-Bonnet quadratic curvature term coupled with a dilaton
\cite{Kanti:1995vq}.
However, these solutions are beyond the scope of the present review.
(A detailed review of the various $d=4$ gravitating solutions with non-Abelian fields was presented a 
decade ago in \cite{Volkov:1998cc}. The case of  $d=4$ asymptotically anti-de Sitter (AdS) solutions which was not covered
in \cite{Volkov:1998cc}, was the subject of the recent review \cite{Winstanley:2008ac}.)
 
These were all static spherically symmetric solutions, some of whose salient properties will be contrasted in their
higher dimensional counterparts to be reported in the next subsection. 
Restricting to solutions with a gauge group $SU(2)$, their basic properties are:

\medskip
\noindent
In the EYM case, the asymptotically flat solutions 
\begin{itemize}
\item
were sphalerons, {\t i.e.} that they were unstable~\cite{Straumann:1990as},
\cite{Bizon:1991hw} since there was no topological charge to supply the
energy with a lower bound, and,
\item
they present radial excitations characterised by a number $k$ of nodes of the function magnetic gauge
 $w(r)$ in (\ref{YMHspheven});
\item
the non-Abelian electric potential necessarily vanishes for both
globally regular and black hole solutions with finite energy \cite{Galtsov:1989ip}, \cite{Bizon:1992pi}.

\end{itemize}

\medskip
\noindent
 
\medskip
\noindent
When a negative cosmological constant is added to the EYM Lagrangian~\cite{Winstanley:1998sn}, \cite{Bjoraker:1999yd},
the asymptotically AdS solutions exhibit new and interesting features, 
 namely that now
\begin{itemize}
\item
the asymptotic value of the function $w(r)$ in (\ref{YMHspheven}) is not fixed {\it a priori}, which leads to
finite mass solutions with a nonvanishing non-Abelian magnetic charge, even without a Higgs field;
\item
stable solutions 
have been shown to exist \cite{Winstanley:1998sn}, \cite{Breitenlohner:2003qj} (this corresponds basically
to the case where the profile of the function $w(r)$ presents no nodes); 
\item
 black holes with non-Abelian hair and a nonspherical topology of the event horizon have been found for  $\Lambda<0$ in
\cite{VanderBij:2001ia}, \cite{Gubser:2008zu};
\item
most importantly in this case, it becomes possible to construct finite energy solutions with
a nonvanishing $A_0$ \cite{Bjoraker:1999yd}, $i.e.$ non-Abelian dyons;
\item
moreover, 
finite mass solutions with AdS asymptotics exist for any $\Lambda<0$.
\end{itemize}

\medskip 

In the case of a positive cosmological constant 
\cite{Torii:1995wv}, \cite{Volkov:1996qj}, \cite{Breitenlohner:2004fp}, \cite{Brihaye:2006kn}, 
 by contrast, 
\begin{itemize}
\item
the EYM solutions with de Sitter (dS) asymptotics exist for sufficiently small
values of $\Lambda$ only;
\item
all solutions have been shown to be unstable, since  $w(r)$ necessarily presents nodes (although
the asymptotic value of the magnetic gauge potential is not fixed {\it a priori});
\item
the electric potential $A_0$ necessarily vanishes for all dS solutions.
\end{itemize}

\medskip
\noindent
In the EYMH (gravitating monopole) case, which differs from the EYM in that a dimensionful constant (the Higgs
vacuum expectation value) appears in the Lagrangian, the solutions with $\Lambda\leq 0$
\begin{itemize}
\item
are topologically stable in the YMH sector, stabilised by the monopole charge;
\item
they present radial excitations characterised by multinode profiles 
in the function $w(r)$ in (\ref{YMHspheven}) as in the YMH
case, and in addition,
\item
due to the presence of the dimensionful constant in the Lagrangian,
they exhibit a Reissner-Nordstr\"om fixed point, which results in the absence of solutions for a range of the
gravitational coupling constant.
\end{itemize}

The picture is more complicated for asymptotically dS gravitating monopole solutions.
Refs. \cite{Brihaye:2005ft}, \cite{Brihaye:2006kn} presented arguments that
\begin{itemize}
\item
although the total mass within the cosmological horizon of the monopoles is finite,
their mass evaluated at timelike infinity generically diverges;
\item
no solutions exist in the absence of a Higgs potential.
\end{itemize}

The $d=4$ asymptotically Minkowski (or AdS) EYM solutions discussed above have axially symmetric generalisations.
The first work in this direction was \cite{Kleihaus:1996vi}, which presented a generalization of the 
Bartnik-McKinnon solutions characterized by a pair of integers $(k,n)$, 
where $n$ is an integer -- the winding number and $k$ is the node number of the amplitude $w(r)$.
The black hole counterparts of these configurations were discussed in \cite{Kleihaus:1997ic}, which 
shows that Israel's theorem \cite{Israel:1967za} does not generalise to
the non-Abelian case ($i.e.$ a static black hole is not necessarily spherically symmetric).
These asymptotically flat solutions were extended afterwards in various directions,
see $e.g.$ \cite{Ibadov:2004rt}, \cite{Kleihaus:2005fs},
\cite{Ibadov:2005rb}, \cite{Kleihaus:2007vf}.
They present also generalizations with a negative cosmological constant, which were discussed in 
\cite{Radu:2001ij}, \cite{Radu:2004gu}, \cite{Mann:2006jc}.
The case of axially symmetric non-Abelian solutions with dS asymptotics was not considered yet in the literature. 

Interestingly, although non-Abelian generalizations of the Kerr-Newmann black hole 
were shown to exist \cite{Kleihaus:2000kg}, \cite{Kleihaus:2002ee},
it turns out that the Bartnik-McKinnon globally regular
solutions admit no asymptotically flat rotating 
generalizations \cite{VanderBij:2001nm}, \cite{Kleihaus:2000kg} 
(however, note that  they were predicted in a perturbative approach \cite{Brodbeck:1997ek}).
  Not completely unexpected, 
spinning EYM solitons were found to exist for AdS asymptotics \cite{Radu:2002rv}, \cite{Mann:2006jc}.

Finally, let us remark that both the EYM and EYMH systems present  nontrivial 
solutions with a NUT charge \cite{Radu:2002hf},  \cite{Brihaye:2005ak}.
These solutions approach asymptotically the Taub-NUT spacetime \cite{NUT} and provide
the non-Abelian counterparts of the U(1) Brill solution \cite{Brill}.
The nonexistence results in \cite{Galtsov:1989ip}, \cite{Bizon:1992pi} are circumvented by these
asymptotically locally flat solutions, which necessarily present a nonzero electric part $A_0$ of the
non-Abelian potential.

\begin{figure}[ht]
\hbox to\linewidth{\hss%
	\resizebox{8cm}{6cm}{\includegraphics{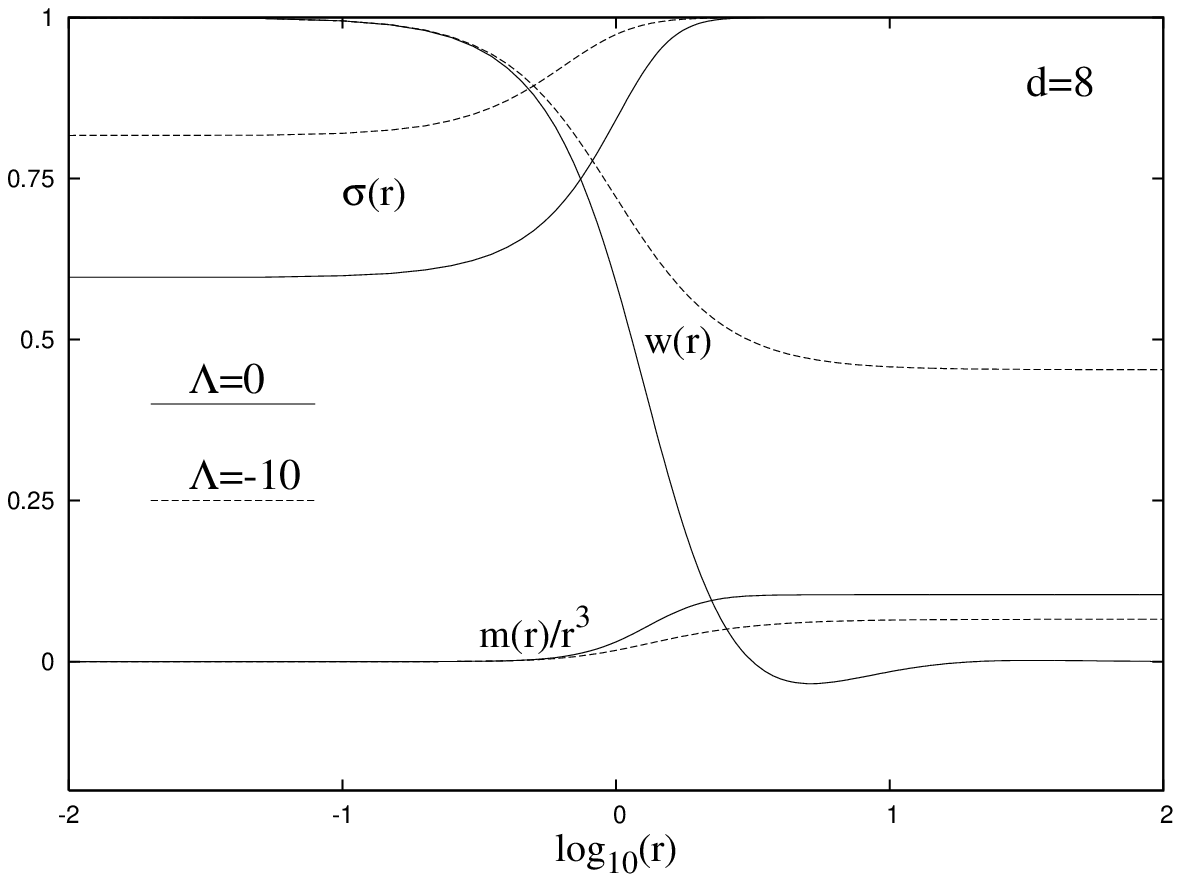}}
\hspace{5mm}%
        \resizebox{8cm}{6cm}{\includegraphics{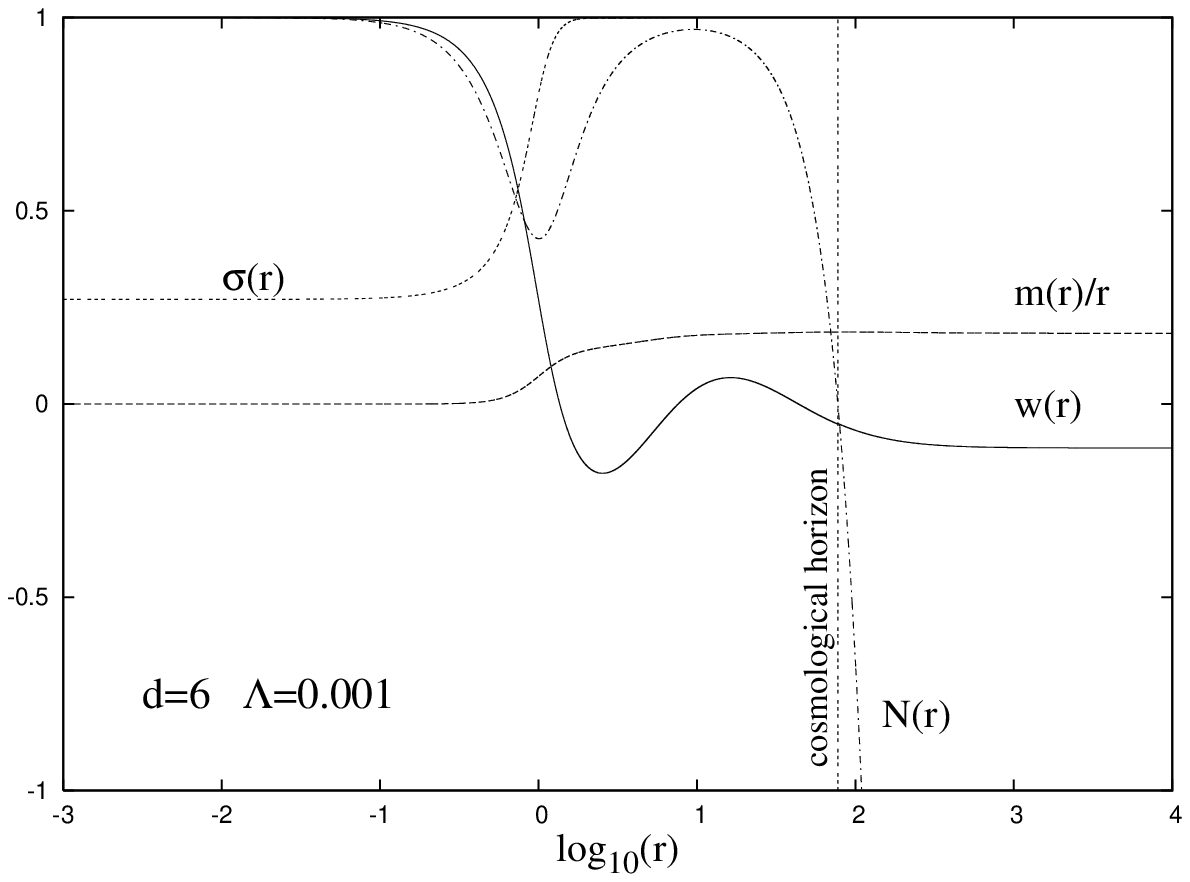}}	
\hss}

\caption{{\small Typical particle-like higher dimensional EYM-$F(2)^2$ solutions in asymptotically flat and anti-de Sitter spacetimes (left) and
de Sitter spacetime (right).
The function  $m(r)$ corresponds to the local mass-energy density.}
 }
\label{fig4}
\end{figure}

\subsubsection{Einstein--Yang-Mills solutions  in higher dimensions}

Gravitating non-Abelian fields in higher dimensions have been considered for the first time
in  \cite{Volkov:2001tb} for $d=5$ 
and a YM model containinig the usual $F(2)^2$ term only. 
For spherically symmetric regular solutions asymptoting to the Minkowski background, it was found  
that their energy  is infinite. Then in Ref. \cite{Okuyama:2002mh} it was proven that the energy of the black hole is
also infinite. Moreover, \cite{Okuyama:2002mh} extended these results to asymptotically AdS solutions.  
When employing the Einstein-Hilbert gravity only, 
one usually defines
\begin{eqnarray}
\label{def-N}
 N(r)=1-\frac{m(r)}{r^{d-3}} -\frac{2\Lambda r^2}{(d-2)(d-1)} ,
 \end{eqnarray}
 the function $m(r)$ being related to the local mass-energy density up to some
$d-$dependent factor.
The results in \cite{Volkov:2001tb},  \cite{Okuyama:2002mh} prove that, in five dimensions, $m(r)\to \log r$, as
$r\to \infty$.

As discussed in \cite{Radu:2005mj}, this is a generic feature of all higher dimensional
EYM solutions with a  $F^2$ term only ($i.e.$ $L_{YM}=\tau_1F_{\mu\nu}^{a}F^{a\mu\nu}$).
Although these configurations are still asymptotically Minkowski, their mass function generically diverges as
$r^{d-5}$ (or as $\log r$ for $d=5$). A similar conclusion is reached when considering \cite{Radu:2005mj} solutions of
a EYM-$\Lambda$ model containinig the usual $F^2$ term only and
aproaching asymptotically an AdS (or dS) background\footnote{Asymptotically AdS solutions with 
diverging mass have been considered by some authors, mainly for a scalar field in the bulk
(see e.g. \cite{Hertog:2004dr}). In this case it might be possible to relax the standard asymptotic
conditions without loosing the original symmetries, but modifying the charges in order to 
take into account the presence of matter fields. A similar approach has been used in ref. \cite{Radu:2005mj}
to assign a finite mass to $d>4$ EYM solutions in a $F^2$ theory with a negative cosmological constant.}.
This can most easily be seen by considering the simplest $w(r)=0$ solution of the EYM equations. This corresponds to
 the gravitating Dirac--Yang
monopoles~\cite{yang,Tchrakian:2008zz}, which are non-Abelian configurations (except in $d=4$ where one has the Abelian Dirac monopole).
These fields are singular at the origin and hence an event horizon should be present.

The result  is a black hole solution, which for $d>5$ has a line element  
\begin{eqnarray}
\label{div-sol}
ds^2=\frac{dr^2}{1-\frac{\mu^2}{r^2}-\frac{M_0}{r^{d-3}} -\frac{2\Lambda r^2}{(d-2)(d-1)}}+r^2 d \Omega_{(d-2)}^2 
-\left(1-\frac{\mu^2}{r^2}-\frac{M_0}{r^{d-3}} -\frac{2\Lambda r^2}{(d-2)(d-1)}\right)dt^2,
\end{eqnarray}
 where $\mu$ is a constant fixed by the YM coupling parameter \cite{Radu:2005mj}, $M_0>0$ is an arbitrary constant
 and $\Lambda$ the cosmological constant.
 This infinite mass configuration generalises to higher dimensions
 the $d=4$ magnetic Reissner-Nordstr\"om black hole and has a number of interesting properties which are discussed in
\cite{Gibbons:2006wd}. The generic solutions of the  $F^2$ EYM model have a more complicated pattern (including
particle-like configuration with a regular origin), but always approach asymptotically the  line element (\ref{div-sol})
(see Figure 1).

\begin{figure}[ht]
\hbox to\linewidth{\hss%
	\resizebox{8cm}{6cm}{\includegraphics{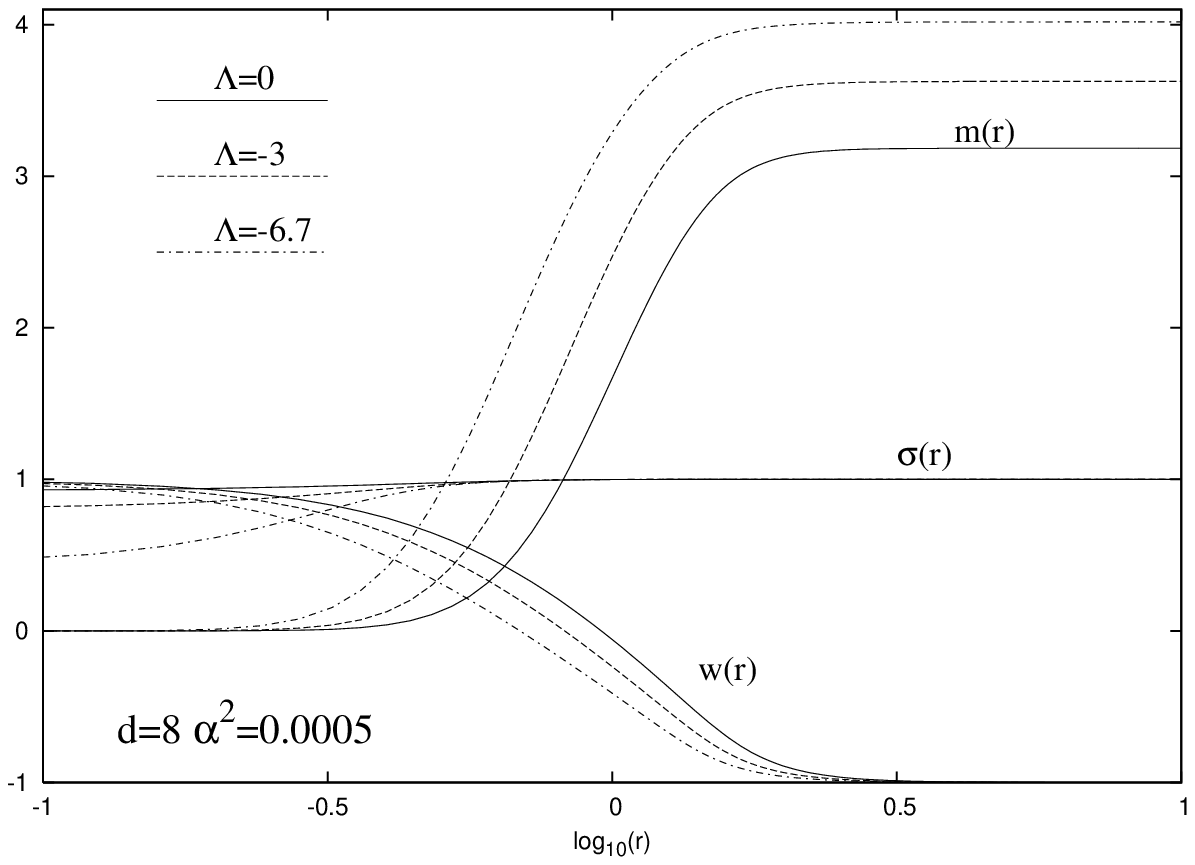}}
\hspace{5mm}%
        \resizebox{8cm}{6cm}{\includegraphics{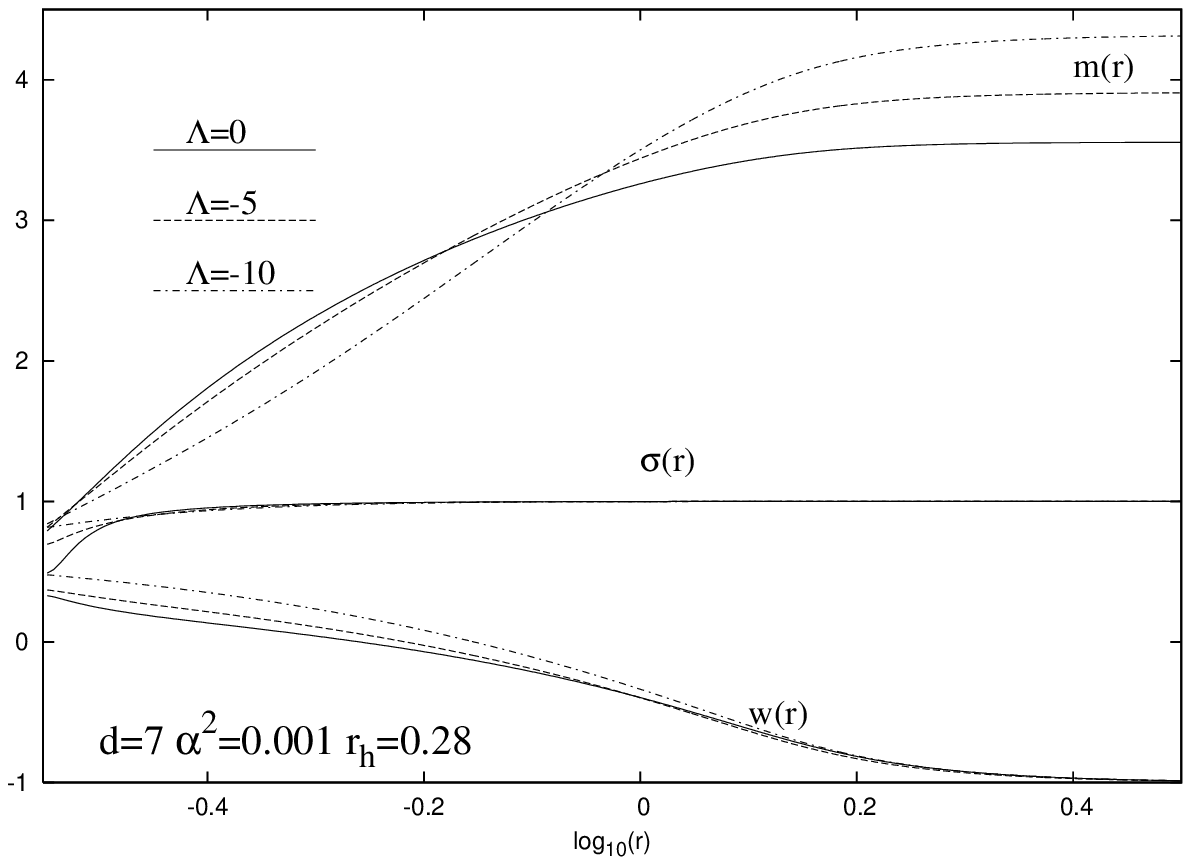}}	
\hss}

\caption{{\small Typical higher dimensional particle-like (left) and 
black hole (right) finite mass solutions of the $p = 1, 2$ EYM theory in asymptotically flat and anti-de Sitter spacetimes.
}
 }
\label{fig4}
\end{figure}

The nonexistence result on finite mass solutions is circumvented by adding the appropriate $p-$YM term(s) to the matter Lagrangian.
As a result, the EYM system presents (at least) one more coupling constant $ \alpha^2 =\sqrt{\tau_1^3 /\kappa_1^2\tau_2}$, 
which usually implies a rich  structure of the solutions. 
Various examples were studied:

\begin{figure}[ht]
\hbox to\linewidth{\hss%
	\resizebox{8cm}{6cm}{\includegraphics{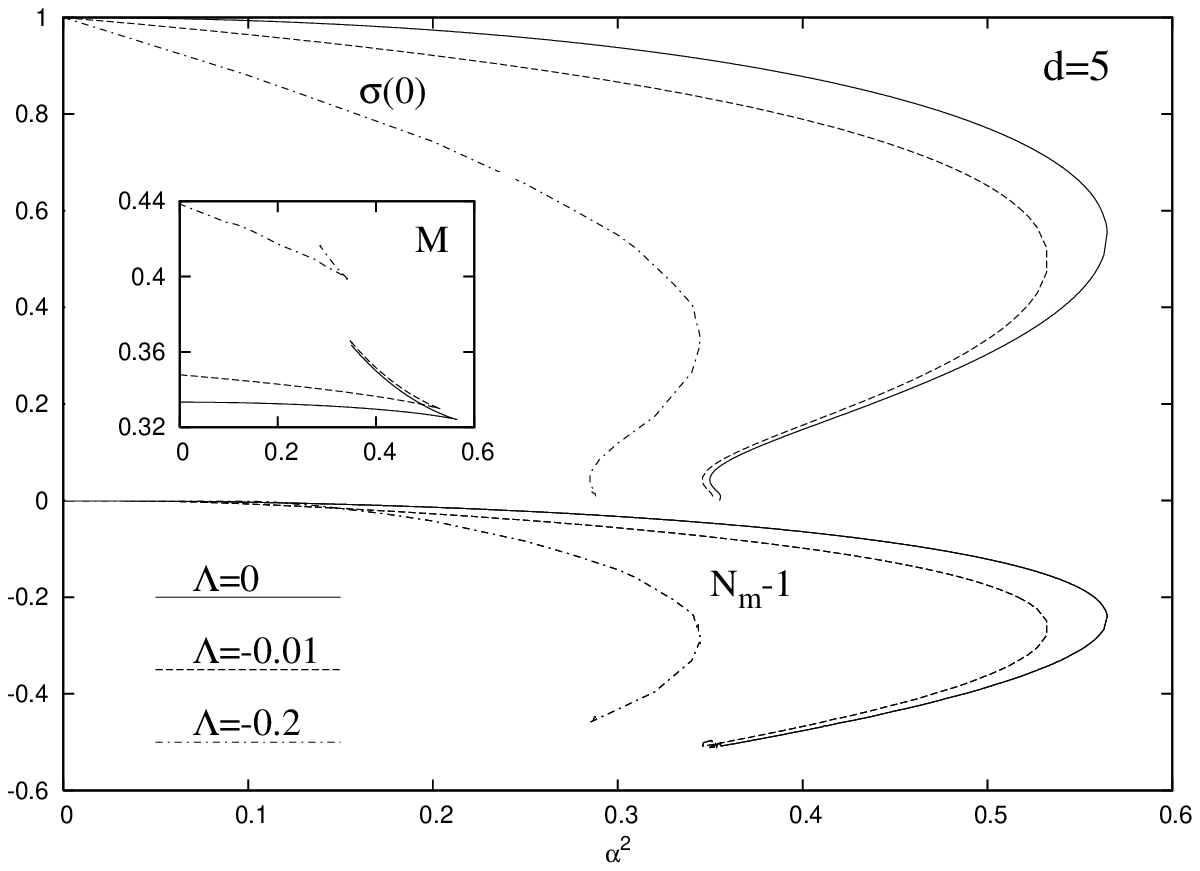}}
\hspace{5mm}%
        \resizebox{8cm}{6cm}{\includegraphics{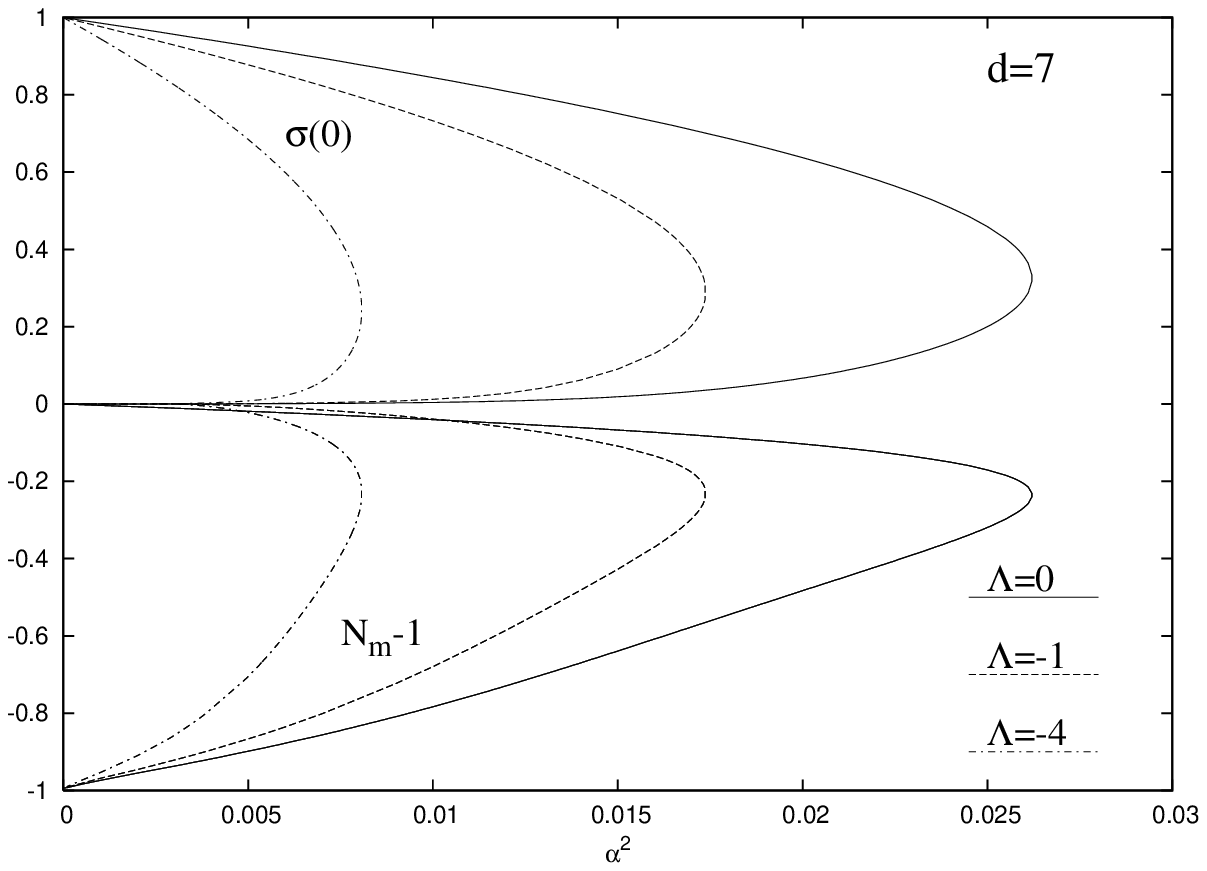}}	
\hss}

\caption{
{\small
The value $N_m$ of the minimum of the metric function $N(r)$,
the mass parameter $M$ as
well as the value of the metric function $\sigma$ at the origin, $\sigma(0)$,
are shown for $d=5$, $d=7$  asymptotically flat and anti-de Sitter particle solutions of the $p = 1, 2$ EYM theory,
as functions of the coupling parameter $\alpha^2 =\sqrt{ \tau_1^3 /\kappa_1^2\tau_2}$ and several values of the cosmological constant 
$\Lambda$. 
}
 }
\label{fig4}
\end{figure}

\medskip
\noindent
\begin{itemize}
\item
in \cite{Brihaye:2002hr} the particle-like solutions of the system consisting of $1-$ and $2-$Einstein-terms 
({\it i.e.}, the Einstein-Gauss-Bonnet system),
$1-$YM and $2-$YM terms in spacetime dimensions $d=6,7,8$, thus exhibiting a dimensionful constant. 
Although a pure gauge configuration is approached in the far field,
these solutions however are not quite direct analogues of the Bartnik-McKinnon solutions because of the
presence of the dimensionful $\tau_2$ constant in the Lagrangian. 
This is analogous with the the gravitating
monopole \cite{Breitenlohner:1991aa}, 
where a dimensionful constant is also involved. 
Unlike the latter, however, there were no radial
excitations in this case.
Moreover, the Gauss-Bonnet in the gravity action does not lead to any new qualitative features of the solutions,

\item
in \cite{Brihaye:2002jg} for particle like solutions of the system
consisting of $1-$Einstein, $1-$YM and  $2-$YM subsystems as above, but in spacetime dimensions $d=5$. 
In addition, in \cite{Brihaye:2002jg}, asymptotically flat black hole solutions are constructed. The fixed point
properties in $d=5$ solutions however differ substantially from those of the $d=6,7,8$ solutions for the same model (see Figure 3).

\item
The fixed point analysis for this model is carried out in \cite{Breitenlohner:2005hx}, where it is found that in addition
to the Reissner-Nordstr\"om fixed point, a new type of fixed point appears. While the Reissner-Nordstr\"om fixed point is
typified by the value of the function $w(r)=0$, the new type of fixed point is typified by the value of the function
$w(r)=1$ and is referred to as a {\it conical fixed point} in \cite{Breitenlohner:2005hx}. It is further shown in
\cite{Breitenlohner:2005hx}, by extending the model judiciously for higher values od $d$ (always keeping only the
$1-$Einstein terms) by higher $p$ YM terms, that this conical singularity appears {\it modulo} every $4p$ dimensions.

\item
EYM systems with negative cosmological constant in higher dimensions are also studied in \cite{Radu:2005mj}. The
finite energy solutions in these models exhibit all the properties seen in 
\cite{Brihaye:2002hr,Brihaye:2002jg,Breitenlohner:2005hx} for $\Lambda=0$.
Different from the $d=4$ case, the higher dimensional AdS solutions necessarily 
have $w(r)\to -1$ as $r\to \infty$, $i.e.$
a pure gauge configuration is approached in the far field (see Figure 2).  
As a consequence, asymptotically AdS  solutions
with negative comological constant cannot support nonvanishing $A_0$ solutions, a {\it per} the argument given in the last
but one item in section {\bf 2.4}.

\item
Higher dimensional EYM systems with positive cosmological constant are studied in \cite{Brihaye:2006xc}. The presence
of a cosmological horizon leads to  a more complicated pattern, where again a conical fixed point appears for $d=5$.
\end{itemize}

\begin{figure}[ht]
\hbox to\linewidth{\hss%
	\resizebox{8cm}{6cm}{\includegraphics{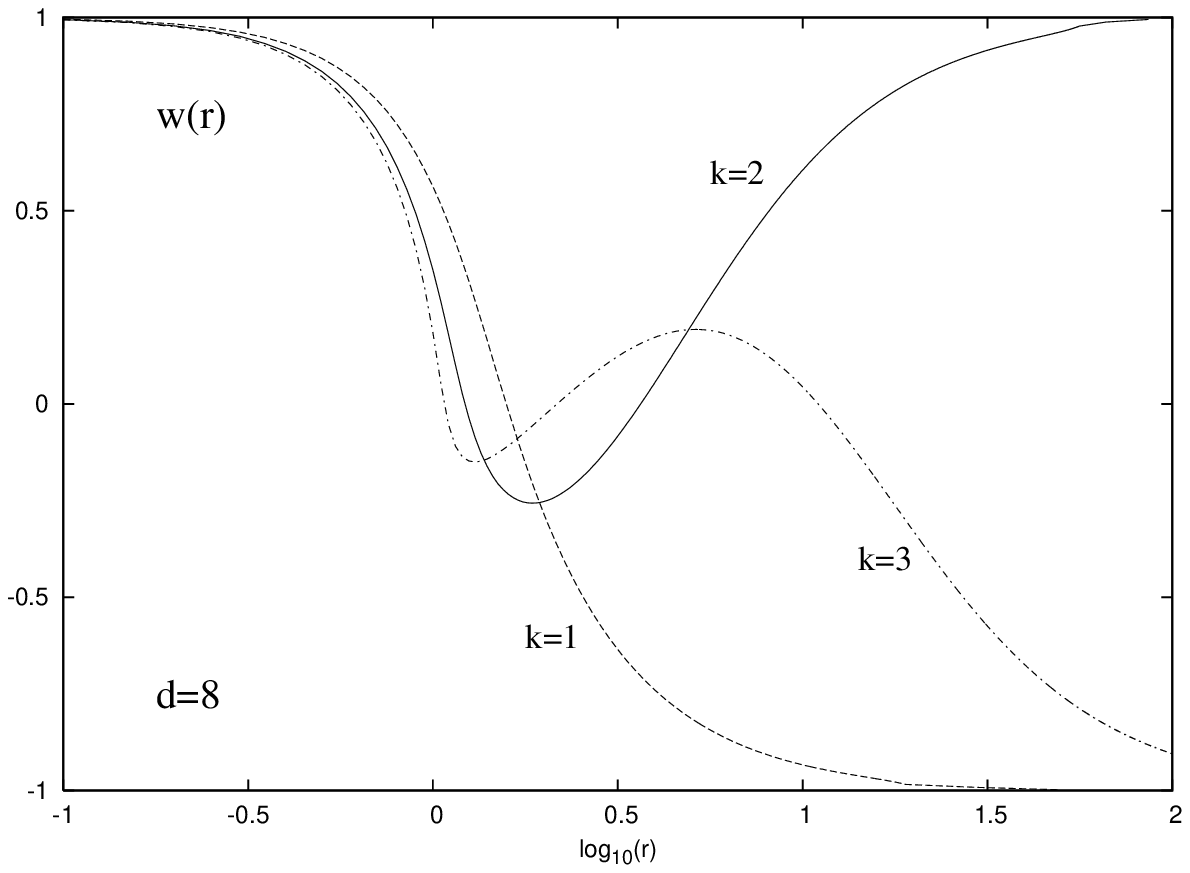}}
\hspace{5mm}%
        \resizebox{8cm}{6cm}{\includegraphics{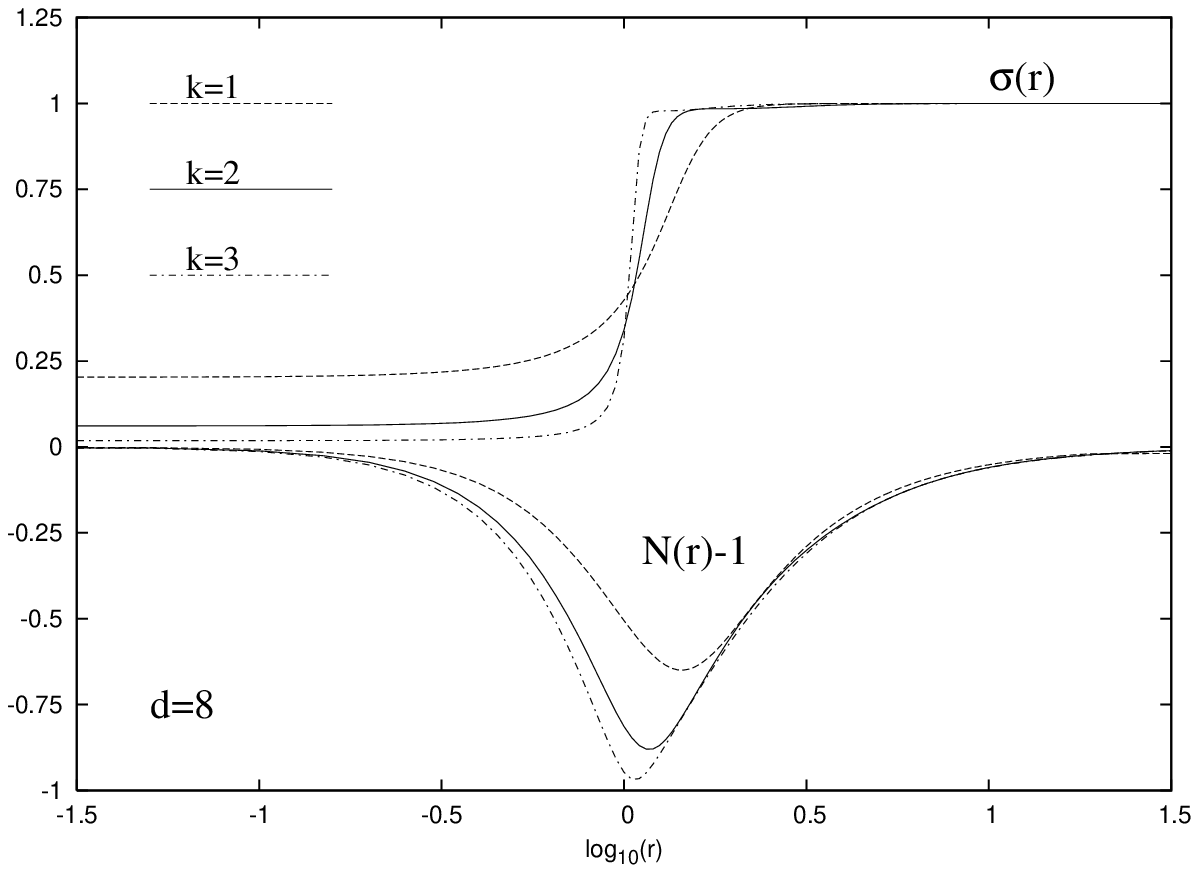}}	
\hss}

\caption{
{\small
The profiles of the metric functions $N(r)$, $\sigma(r)$ and gauge function $w(r)$ are presented for $k-$node
globally regular solutions of the $p = 2$ gravity-Yang-Mills model in $d = 8$ dimensions.
}
 }
\label{fig4}
\end{figure}


All the above listed EYM solutions pertain to models motivated entirely by the criterion of satisfying the scaling
requirement for finite energy. In a further work~\cite{Radu:2006mb} in $d=4p$,
the models were chosen according to the  criterion that only $p-$Einstein and $p-$YM terms appear in the
Lagrangian\footnote{The ref. \cite{Radu:2006mb} presents also an exact solution
for the $p$-th Einstein-Yang-Mills system in $d=2p+1$ dimensions.}.
In these cases no dimensionful constant appears
in the Lagrangian and the properties of the solutions
 are entirely similar to those of the Bartnik-McKinnon solution (in particular the existence of radial excitations),
except that qualitative features are appreciably magnified with increasing $p$ (see Figure 4).

There is also the question of non--spherically symmetric EYM solutions in higher dimensions. Since all EYM
solutions are constructed numerically, the problem here is to relax this symmetry such that the numerical process
remains tractable. The most s traightforward step would be the imposition of {\it axial} symmetry in the $D-$spacelike
dimensions, {\it i.e.}, by imposing spherical symmetry in the $(D-1)-$dimensional subspace, thus reducing the problem
to a $2-$dimensional PDE. This can be done readily for
arbitrary $d=D+1$, but unfortunately the implementation of the numerical integration becomes problematic when removing
the gauge arbitrariness. Instead, for $d=5$,
a system of $2-$dimensional PDE's can be obtained when an azimuthal symmetry is imposed
in each of the two planes of the $4-$dimensional $t=const.$ spacelike subspace. 
In principle, this can be generalized for any odd, $d=2n+1$, dimensional
spacetime, and then the reduced problem will be that of a $n-$dimensional PDE's. This limits one to the bi-azimuthal
regime in $d=4+1$, for practical reasons. 

Other than this static result, there are two other indirect $d>4$ results in the
literature which are not spherically symmetric. One is the case where there is a rotation in the two spacelike
sub-planes in the $4+1$ dimensional case, and the other concerns a rather different topology of the spacetime.
 
\begin{itemize}
\item
Ref.~\cite{Radu:2007jb} discussed static solutions in $d=4+1$ dimensions
of a EYM system with  bi-azimuthal symmetry in four spacelike dimensions.
They generalise the  configurations in  \cite{Brihaye:2002hr}, \cite{Brihaye:2002jg}, both particle-like and black hole
solutions being found to exist.
It is interesting that the fixed point structures discovered in the spherically
symmetric cases in \cite{Brihaye:2002hr,Brihaye:2002jg,Breitenlohner:2005hx,Radu:2005mj,Brihaye:2006xc} in $d=5$,
manifest themselves for these bi-azimuthally symmetric solutions, although a rigorous fixed point analysis like in
\cite{Breitenlohner:2005hx} is not analytically accessible in this case.

\item
$d=5$ rotating EYM black hole solutions in the usual EYM model ($i.e.$,  with a  $F^2$ term only)
with negative cosmological constant were constructed in \cite{Brihaye:2007tw}.  The rotation in question was that of
two equal angular momenta in the two spacelike 2-planes, which is known to reduce the $2-$dimensional PDE
problem to a $1-$dimensional ODE one   ($i.e.$ the angular
dependence is factorised in the ansatz). In this respect, it is a system defined by a single radial variable, but does not
not describe a spherically symmetric field configuration.
 Different from the static case, no spinning regular solution 
is found for a vanishing event horizon radius.
As expected,  the mass of these solutions as defined in the usual way, diverges. 
However, a finite mass  can be assigned by using a suitable version
of the bounday counterterm regularization method \cite{Balasubramanian:1999re}. (We would of course expect finite
energy solutions in this case too, had the $F^4$ YM term been included.)

\item
In AdS spacetime the topology of the horizon of a black hole solution is no
longer restricted to be $spherical$.   It can be $planar$, or, $hyperbolic$ instead. 
A surprising result reported in \cite{Manvelyan:2008sv} is that, for $\Lambda<0$,
there are $d>4$ asymptotically AdS, finite mass black hole solutions with a $planar$
topology of the event horizon, even in a theory without higher derivative terms in the YM curvature. This contrasts
with the corresponding black holes with a $spherical$ topology of the event horizon, which have infinite mass
\cite{Radu:2005mj}. The case of a $hyperbolic$ topology of the horizon has not been considered yet in the
literature for $d>4$. As in the previous example, the (consistent) Ansatz used for the YM field does reduce the
PDE's to a system of one dimensional ODE's in terms of a radial variable, but likewise does not describe a spherically
symmetric field configuration.
 
\end{itemize}

\begin{figure}[ht]
\vspace{-3.cm}
\hbox to\linewidth{\hss%
	\resizebox{8cm}{9cm}{\includegraphics{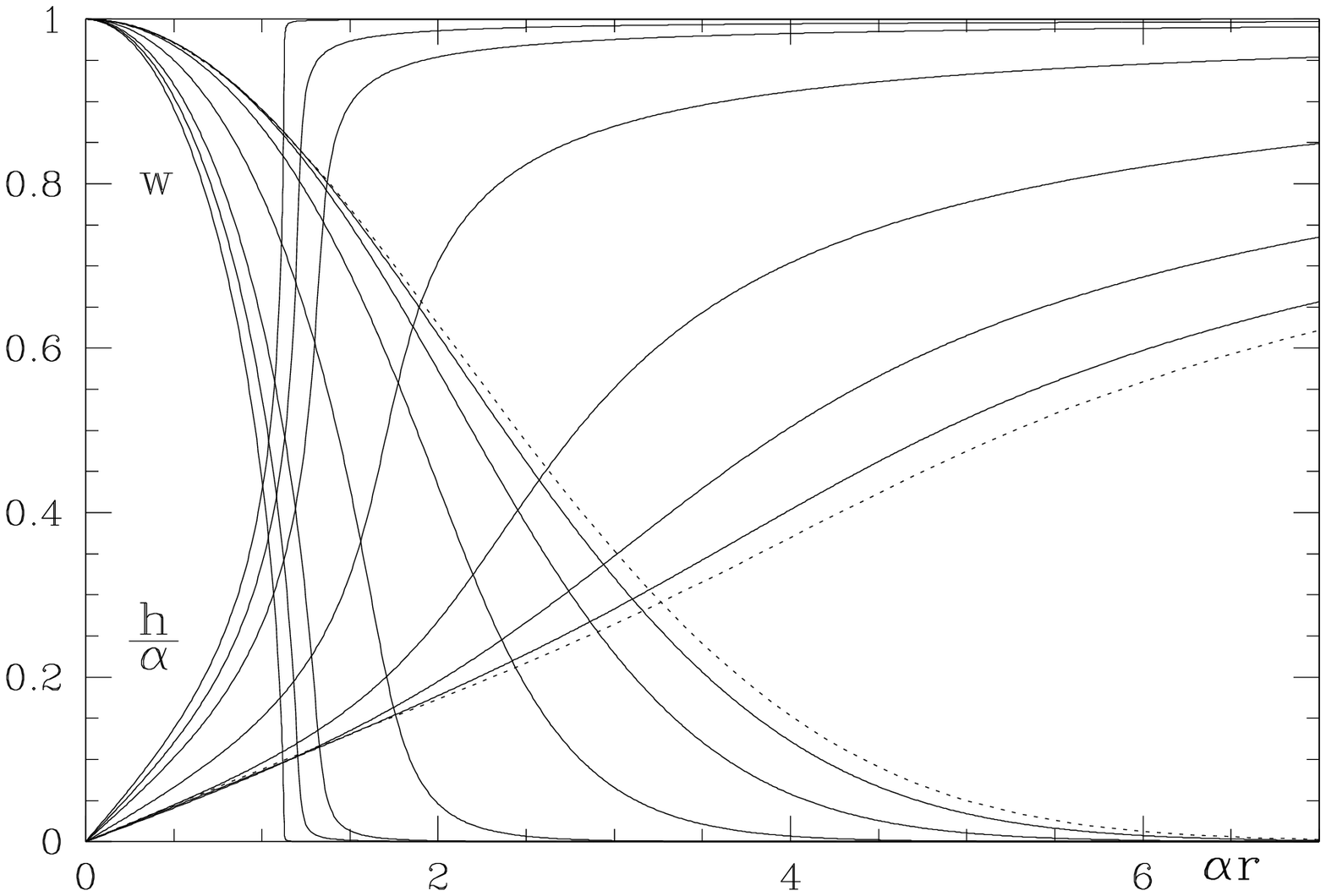}}
\hspace{5mm}%
        \resizebox{8cm}{9cm}{\includegraphics{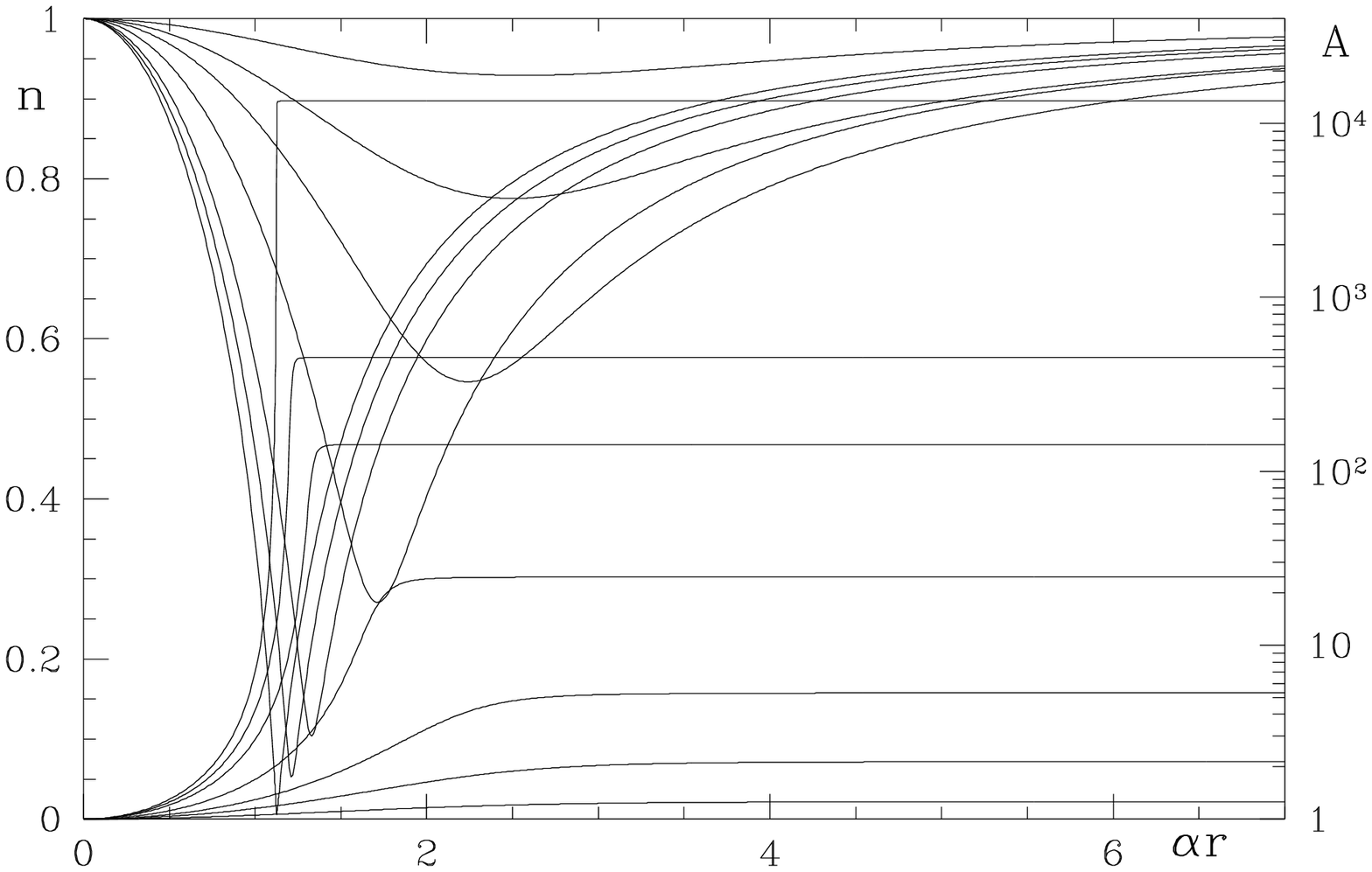}}	
\hss}
\caption{
{\small
Fundamental solutions corresponding to several values of the coupling constant $\alpha=\eta(\tau_p/\kappa_p)^{1/2p}$ 
are shown 
for  $p=4$ higher dimensional 
gravitating YMH monopoles in \cite{Breitenlohner:2009zi}; the dotted curves are for the monopole in flat space. }
}
\label{fig4}
\end{figure}
 
 In addition to the higher dimensional gravitating YM fields described above, there has been some work also 
studying gravitating monopoles in higher dimensions. Recently 
a very particular hierarchy of selfgravitating YMH models in $4p$ dimensions was studied in detail
in \cite{Breitenlohner:2009zi}. This family of models,
whose flat space monopoles were constructed in \cite{Radu:2005rf}, is the most direct generalisation of the
$d=3+1$ Georgi-Glashow
model (in the BPS limit), and the only one for which the Bogomol'nyi inequalities can be saturated. 
Both  regular and black hole solutions have been 
constructed in \cite{Breitenlohner:2009zi}, which exhibit all the generic properties of the well known $d=4$ gravitating monopoles (the profiles
of typical solutions are exhibited in Figure 5).  In
higher dimensions, there are many other types of monopoles, $e.g.$, that in $d=4+1$ in \cite{O'Brien:1988xr}, and that in
$d=3+1$ in \cite{Kleihaus:1998gy}. The selfgravitating versions of these are not studied to date. 

Finally, let us mention the case of $d>4$ non-Abelian solutions 
with codimensions\footnote{A detailed review of these solutions is presented in 
\cite{Volkov:2006xt}, \cite{Hartmann:2008hq}.}.
As mentioned already, the situation of $d=5$ with one codimension is the only case discussed
in a systematic way in the literature, mainly for a metric ansatz 
which is spherically symmetric in a four dimensional perspective
\begin{eqnarray}
\label{metrica}
ds^2 = e^{- 2\phi(r)/\sqrt{3}}\left(\frac{dr^2}{N(r)}
+ r^2d\Omega_{(2)}^2- N(r) \sigma^2(r) dt^2\right)
 + e^{ 4\phi(r)/\sqrt{3}}(dx^5)^2,
\end{eqnarray}
$x^5$ being the extra-direction and $\phi(r)$ corresponding to a dilaton field.

\begin{itemize}
\item
The KK theory  possesses in this case a variety of 
interesting non-Abelian configurations, including axially symmetric generalizations
\cite{Volkov:2001tb}, \cite{Okuyama:2002mh}, \cite{Brihaye:2004kh}, \cite{Hartmann:2004tx}.
After performing a KK reduction, they correspond to $d=4$ particle like and black hole 
solutions in a Einstein-Yang-Mills-Higgs-U(1)-dilaton theory  \cite{Brihaye:2005pz}.
\item
EYM black strings and vortices with a cosmological constant were discussed in ref. \cite{Brihaye:2007jua}.
\item
The inclusion of of higher order terms of the
YM curvature is optional for $d=5$ black strings and vortices, since they possess a finite mass per unit lenght
of the extradimension already in the usual $F^2$ theory. In this case, the higher derivative terms do not
affect the basic properties of the solutions.
\end{itemize}

\subsection{EYM solutions with Euclidean signature}
 

While this review concerns primarily fully gravitating EYM and EYMH solutions in Lorentzian signature, it is
reasonable to allude to EYM solutions in Euclidean signature, especially since these were the first such solutions
that appeared in the literature. Such solutions, including those on (Euclidean)
Schwarzschild and de Sitter~\cite{Charap:1977re,Charap:1977ww,BoutalebJoutei:1979va} and
Taub-NUT~\cite{Pope:1978kj} backgrounds have been studied long ago. 
More recently there have been further
investigations~\cite{Bianchi:1996zj,Etesi:2002cc,Cherkis:2009jm} of YM fields on fixed backgrounds.

 All of the known EYM fields with $A_0\neq 0$ and Euclidean signature in the literature are given on fixed gravitational backgrounds.
None with gravity backreacting on the YM fields is known, which puts these solutions on a different footing to those
with Lorenzian signature duscussed above.

Starting with backgrounds which are analytic continuations of relevant solutions with
Lorenzian signatures, we note that
\begin{itemize}
\item
The exact solutions of Charap and
Duff~\cite{Charap:1977re,Charap:1977ww}, as also their higher dimensional counterparts~\cite{Radu:2007az}, are by
construction given on gravitational backgrounds for which the the $2p-$form Riemann curvature is double--self-dual.
Such metrics satisfy the hierarchy of vacuum Einstein equations (with or withour cosmological constant) so that by
construction, these are EYM solutions on backgrounds of fixed curvature.

\item
In an effort to go away from fixed backgrounds, a direct numerical method was employed in \cite{Brihaye:2006nk},
using a (Euclidean) Schwarzschild metric and static YM fields in $d=4$. But being a static field configuration, the 'electric' YM
potential $A_0$ assumed the role of a Higgs field and the resulting solutions turned out to be self-dual 'deformed
Prasad-Sommerfield monopoles', again on a fixed background. 
These solutions are different from those in \cite{Charap:1977re,Charap:1977ww}, as shown $e.g.$
by a computation of their action.  The higher dimensional analogues of this type of solutions in $d=4p$
were constructed in \cite{Radu:2007az}. 

\item
 In a further development  beyond \cite{Brihaye:2006nk}, a number of other static spherically symmetric $d=4$ 
 metric backgrounds
 were employed in \cite{Brihaye:2006bk} to construct Euclidean non-Abelian solutions. 
 All these resulted in
selfdual YM solutions, on the basis of which it was conjectured that  for  any $d=4$ (Euclidean) static spherically symmetric
metric, the solutions satisfy the $d=4$ Yang-Mills self-duality equations.
An analytic proof of this conjecture has been given in \cite{Radu:2007az}, where
the $4p-$dimensional analogues of these were also constructed, satisfying the self-duality equations
(\ref{sd-4p}). 
 
The solutions described in the above three items generalise 
both the Charap-Duff \cite{Charap:1977re,Charap:1977ww}
and the solutions in \cite{Brihaye:2006bk} in $d=4$, to $d=4p$, the properties of the four dimensional case
being generic. 

\item
Another property of the known gravitating instantons which is related to the fact they are given on fixed backgrounds
is, that they are always (Euclidean) time independent. Even when an explicit time dependence is buit in to the YM
Ansatz, it turns out that the solutions are either time independen~\cite{Tekin:2002mt}, or in the presence of a
cosmological constant, that the Pontryagin charge of the instanton is noninteger~\cite{Sarioglu:2009du}. We believe that
this is a result of having used a static metric. Relaxing this last property may be interesting but promises to lead to
a nontrivial numerical problem.

\end{itemize}

 The metric Ansatz (\ref{metric}) in all above described Euclidean EYM fields, makes a distinction between the (Euclidean) 
 time and the space
coordinates.  A different type of solutions were found in another setting,  the metric Ansatz being
spherically symmetric  in $d$ dimensions 
\cite{Maldacena:2004rf,Radu:2007az}  
\be
\label{metric2}
ds^2=d\rho^2+f^2(\rho)d \Omega_{(d-1)}^2,
\ee
where $f(\rho)$ is a function fixed by the gravity-matter field equations, 
$\rho$ being the radial coordinate,  $\rho=\sqrt{|x_{\mu}|^2}$ and $\hat x_{\mu}=x_{\mu}/\rho$ is 
the unit radius vector.
The YM ansatz compatible with the symmetries of the above line element is 
\be
\label{YMsph2}
A_{\mu}=\left(\frac{1-w(\rho)}{\rho}\right)\,\Sigma_{\mu\nu}^{(\pm)}\hat x_{\nu}\,,
\ee
where the spin matrices are precisely those used in (\ref{YMHsphodd}), (\ref{YMHspheven}).

The resulting reduced one dimensional YM Lagrangian for the $p$-th term in the YM hierarchy
read
\bea
L^{(p,d)}_{\rm{YM}}&=&\frac{\tau_p}{2\cdot (2p)!}
\frac{(d-1)!}{(d-2p)!} 
f^{d-4p+1} (w^2-1)^{2p-2}\left(w'^2+\frac{d-2p}{2p}\frac{(w^2-1)^2}{f^2}\right)~.
\label{LYMpd-s2}
\eea
For any choice of the metric function $f(\rho)$, the solution of the 
YM self-duality equation (\ref{sd-4p})  in $d=4p$ dimensions  reads
\bea
w(\rho)=\frac{1-c_0e^{\mp 2\int \frac{d\rho}{f(\rho)}}}{1+c_0e^{\mp 2\int \frac{d\rho}{f(\rho)}}}~,
\label{eq-s3}
\eea
where $c_0$ is an arbitrary positive constant.
As discussed in \cite{Radu:2007az}, the action of these non-Abelian solutions is finite for any value of $p$.
For $f(\rho)=\rho$ one recovers the $d=4p$ generalisation of the
BPST instanton first found in \cite{Tchrakian:1984gq}, with $w=(\rho^2-c)/(\rho^2+c)$.
An AdS background $f(\rho)=\rho_0\sinh \rho/\rho_0$ leads to a 
$d=4p$ generalisation of the $d=4$ AdS selfdual instantons in \cite{Maldacena:2004rf}, with
$w=(\tanh^2(\rho/2\rho_0)-c)/(\tanh^2(\rho/2\rho_0)+c)$.
The $d=4p$ selfdual instantons on a sphere (euclideanised dS space)
are found by taking $\rho_0\to i \rho_0$ in the corresponding AdS relations.
 
 As selfdual solutions, these are also fixed background YM fields. Moreover, on curved backgrounds the
selfduality equation saturating the inequality \re{ineq} can be solved.
 Thus in
Ref. \cite{Radu:2007az} systems consisting of the superposition of two members of the
YM hierarchy, say those labeled by $p$ and $q$, with $d=2(p+q)$ were considered as well.
Solutions of such selfduality equations were also discussed in \cite{Kihara:2007di}, \cite{Kihara:2007vz}.

\section{Summary and outlook}

We have reviewed a number of results on  Einstein--Yang-Mills (EYM) solutions, with special
emphasis on the new features that arise for spacetime dimensions $d> 4$, mainly in Lorentzian signature.
The EYM solutions that we have considered are those for fully backreacting gravity with matter.
It turns out that these solutions are all constructed numerically and no relevant closed form solutions are known.
As such, the question of imposition of symmetries with the aim of reducing the dimensionality of the Euler--Lagrange
equations becomes a very important feature of these investigations, at least as important as in the case of solutions
that can be expressed in closed form.

  Higher dimensional EYM fields with Euclidean signature are also mentioned, but only in passing since they are
not on the same footing as their Minkowskian counterparts. They are exclusively YM fields on fixed backgrounds, in
all dimensions. 

A  salient feature of higher dimensional non-Abelian solutions 
is that in all $d=D+1$ dimensions, for $D\ge 4$  the usual EYM system cannot
support asymptotically flat, finite energy solutions. This is because of the inappropriate scaling properties of that
system and is remedied by the addition of higher order curvature terms. The higher order terms in question are
exculsively ones that are constructed with higher order YM curvature forms, which in our nomenclature are the
$p-$YM members of the YM hierarchy, the $1-$YM being the usual YM system. The resulting YM equations contain no higher
derivatives of the gauge potential than second. The higher order gravitational curvature systems
($e.g.$, Gauss-Bonnet and higher) are not possible to exploit for this purpose. Higher order gravities have nonetheless
been employed in some contexts, not out of necessity, but for emphasising qualitative features of certain EYM and
EYM-Higgs (EYMH) solutions, which get magnified if dimensions of the consituent terms in the system are suitably
matched (see e.g. \cite{Radu:2006mb}).

 The necessity of employing higher curvature members of the YM hierarchy to enable finite energy holds both for
asymptoticall flat EYM, as well as in the presence of a cosmological constant. There is however an exception to this rule,
namely in the case of asymptotically AdS EYM black hole solutions with a Ricci flat horizon geometry. 
In that case, which is of particular interest for applications to AdS/CFT, it turns out that the usual EYM system can
support finite energy solutions in all dimensions~\cite{Manvelyan:2008sv}. There, the nonexistence proof for solutions
with nonvanishing 'electric' YM potential does not hold.

Concerning the physical justification of employing higher order YM and Riemann curvature terms, one notes that these
occur in the low energy effective action of string theory~\cite{GSW,Pol}. Indeed there is some controversy on the precise
structure of such terms, especially in the YM case~\cite{Tseytlin,BRS,CNT}, but we do not take account of these
considerations here. From the point of view of applications for higher dimensional EYM, these can be used to extend the
results of \cite{Gauntlett:1992nn,Bergshoeff:2006bs,Gibbons:1993xt} to higher dimensions. Some of these authors, and
\cite{Minasian:2001ib,Polchinski:2005bg}, employ only solutions in closed form, but further developments would
necessitate the fully gravitating solutions which are evaluated only numerically. In any case, our focus here was
exclusively on the existence and the generic properties of EYM
solutions in higher dimensions, rather than their physical applications.

 Nearly all gravitating non-Abelian solutions in $d=D+1$ dimensions with $D\ge 4$ reviewed in this work are static and
spherically symmetric. This contrasts with the situation in $3+1$ spacetime dimensions where axial symmetry in $3$ space
dimensions is really $azimuthal$ symmetry and the non-Abelian field configurations are encoded with a vortex (winding)
number. The latter turns out to be an essential tool in the numerical constructions
\cite{Kleihaus:1996vi,Kleihaus:1997ic}. In higher than four spacetime dimensions however, axial symmetry implies
spherical symmetry in one dimension lower than $D$ and there is no winding number associated with the axially symmetric
fields. 
However, the removal of the gauge arbitrariness turns out to be a much harder
problem in this case, presenting a technical obstacle. Thus, the only non-spherically symmetric EYM solutions
studied to date are ones in $4+1$ spacetime with bi-azimuthal symmetry, when there are two vortex numbers encoding the
symmetries of the field configurations~\cite{Radu:2007jb}.

Looking further ahead we should note that in recent years it has become clear that as the dimension $d$ increases,
the phase structure of the (non-spherically symmetric) solutions of the Einstein equations
becomes increasingly intricate and diverse, already in the vacuum case
(see $e.g.$ the recent work \cite{Emparan:2009cs}).
It is very likely that, given the interplay in this case between internal group symmetries and
the spacetime symmetries, the extension of known such vacuum solutions to a non-Abelian
matter content would lead to a variety of new unexpected configurations.
This is a promising but technically difficult direction for the future.

Finally, we mention the higher dimensional Euclidean EYM fields, which is of marginal interest here since none of the
known such solutions are genuinely selfgravitating, but rather YM fields on fixed backgrounds. In all dimensions,
including four, these turn out to be selfdual YM fields, and as such are restricted to even dimensions only. Another
aspect of this restriction turns out to be that these YM fields appear to be (Euclidean-) time independent and are not YM
instantons at all. It would be interesting to construct non-selfdual Euclidean EYM fields and inquire whether these,
if they exist, describe genuine time dependent instantons.

\medskip
\medskip
\noindent
{\bf Acknowledgements}
\\
This material was presented at the Heraeus School in Bremen, September 2008. We thank the organisers, Jutta Kunz and
Claus Laemmerzahl for giving us this opportunity. We have greatly benefited from discussions and collaboration with
Peter Breitenlohner, Yves Brihaye, Amithabha Chakrabarti, Betti Hartmann, Jutta Kunz, Burkhard Kleihaus, Dieter Maison, Yasha Shnir,
Michael Volkov and Yisong Yang. This work is carried out in the framework of Science Foundation Ireland (SFI) project
RFP07-330PHY. The work of ER was supported by a fellowship from the Alexander von Humboldt Foundation.

\begin{small}

\end{small}


\begin{thebibliography}{99}
\bibitem{Bartnik:1988am}
  R.~Bartnik and J.~Mckinnon,
  Phys.\ Rev.\ Lett.\  {\bf 61} (1988) 141.
\bibitem{Volkov:1998cc}
  M.~S.~Volkov and D.~V.~Gal'tsov,
  Phys.\ Rept.\  {\bf 319} (1999) 1
  [arXiv:hep-th/9810070].
\bibitem{gravskyrme}  
H. Luckock, {\it Black hole skyrmions}, in H.J. De Vega and N. Sanches (editors),
{\it "String Theory, Quantum Cosmology and Quantum Gravity, Integrable and
Conformal Integrable Theories"}, page 455. World Scientific, 1987.    
\bibitem{Volkov:1989fi}
  M.~S.~Volkov and D.~V.~Galtsov,
  JETP Lett.\  {\bf 50} (1989) 346
  [Pisma Zh.\ Eksp.\ Teor.\ Fiz.\  {\bf 50} (1989) 312].
\bibitem{Kuenzle:1990is}
  H.~P.~Kuenzle and A.~K.~M.~Masood- ul- Alam,
  J.\ Math.\ Phys.\  {\bf 31} (1990) 928.
\bibitem{Bizon:1990sr}
  P.~Bizon,
  Phys.\ Rev.\ Lett.\  {\bf 64} (1990) 2844.
\bibitem{Breitenlohner:1991aa}
  P.~Breitenlohner, P.~Forgacs and D.~Maison,
  Nucl.\ Phys.\  B {\bf 383} (1992) 357.
\bibitem{Breitenlohner:1994di}
  P.~Breitenlohner, P.~Forgacs and D.~Maison,
  Nucl.\ Phys.\  B {\bf 442} (1995) 126
  [arXiv:gr-qc/9412039].
\bibitem{Lee:1991vy}
  K.~M.~Lee, V.~P.~Nair and E.~J.~Weinberg,
  Phys.\ Rev.\  D {\bf 45} (1992) 2751
  [arXiv:hep-th/9112008].   
\bibitem{Winstanley:1998sn}
  E.~Winstanley,
  Class.\ Quant.\ Grav.\  {\bf 16} (1999) 1963
  [arXiv:gr-qc/9812064].
\bibitem{Bjoraker:1999yd}
  J.~Bjoraker and Y.~Hosotani,
  Phys.\ Rev.\ Lett.\  {\bf 84} (2000) 1853
  [arXiv:gr-qc/9906091].
\bibitem{Bjoraker:2000qd}
  J.~Bjoraker and Y.~Hosotani,
  Phys.\ Rev.\  D {\bf 62} (2000) 043513
  [arXiv:hep-th/0002098].
\bibitem{Volkov:1996qj}
  M.~S.~Volkov, N.~Straumann, G.~V.~Lavrelashvili, M.~Heusler and
O.~Brodbeck,
  Phys.\ Rev.\  D {\bf 54} (1996) 7243
  [arXiv:hep-th/9605089].
\bibitem{Volkov:2001tb}
  M.~S.~Volkov,
  Phys.\ Lett.\  B {\bf 524} (2002) 369
  [arXiv:hep-th/0103038].
\bibitem{Okuyama:2002mh}
  N.~Okuyama and K.~i.~Maeda,
  Phys.\ Rev.\  D {\bf 67} (2003) 104012
  [arXiv:gr-qc/0212022].
\bibitem{GSW}
M.B. Green, J.H. Schwarz and E. Witten, {\it Superstring Theory},
Cambridge University Press, Cambridge, 1987.
\bibitem{Pol}
J. Polchinski, {\it TASI lectures on D-branes}, hep-th/9611050.     
\bibitem{Brihaye:2002hr}
  Y.~Brihaye, A.~Chakrabarti and D.~H.~Tchrakian,
  Class.\ Quant.\ Grav.\  {\bf 20} (2003) 2765
  [arXiv:hep-th/0202141].
\bibitem{Breitenlohner:2005hx}
  P.~Breitenlohner, D.~Maison and D.~H.~Tchrakian,
  Class.\ Quant.\ Grav.\  {\bf 22} (2005) 5201
  [arXiv:gr-qc/0508027].
  
\bibitem{Brihaye:2002jg}
  Y.~Brihaye, A.~Chakrabarti, B.~Hartmann and D.~H.~Tchrakian,
  Phys.\ Lett.\  B {\bf 561} (2003) 161
\bibitem{Radu:2005mj}
  E.~Radu and D.~H.~Tchrakian,
  Phys.\ Rev.\  D {\bf 73} (2006) 024006
  [arXiv:gr-qc/0508033].
\bibitem{Brihaye:2006xc}
  Y.~Brihaye, E.~Radu and D.~H.~Tchrakian,
  Phys.\ Rev.\  D {\bf 75} (2007) 024022
  [arXiv:gr-qc/0610087].
\bibitem{Radu:2006mb}
  E.~Radu, C.~Stelea and D.~H.~Tchrakian,
  Phys.\ Rev.\  D {\bf 73} (2006) 084015
  [arXiv:gr-qc/0601098].
\bibitem{Breitenlohner:2009zi}
  P.~Breitenlohner and D.~H.~Tchrakian,
  arXiv:0903.3505 [gr-qc].
\bibitem{Kleihaus:1996vi}
  B.~Kleihaus and J.~Kunz,
  Phys.\ Rev.\ Lett.\  {\bf 78} (1997) 2527
  [arXiv:hep-th/9612101].
\bibitem{Kleihaus:1997ic}
  B.~Kleihaus and J.~Kunz,
  Phys.\ Rev.\ Lett.\  {\bf 79} (1997) 1595
  [arXiv:gr-qc/9704060].
\bibitem{Hartmann:2000gx}
  B.~Hartmann, B.~Kleihaus and J.~Kunz,
  Phys.\ Rev.\ Lett.\  {\bf 86} (2001) 1422
  [arXiv:hep-th/0009195].
\bibitem{Radu:2007jb}
  E.~Radu, Y.~Shnir and D.~H.~Tchrakian,
  Phys.\ Lett.\  B {\bf 657} (2007) 246
  [arXiv:0705.3608 [hep-th]].
\bibitem{Charap:1977re}
  J.~M.~Charap and M.~J.~Duff,
  Phys.\ Lett.\  B {\bf 71} (1977) 219.
\bibitem{Charap:1977ww}
  J.~M.~Charap and M.~J.~Duff,
  Phys.\ Lett.\  B {\bf 69} (1977) 445.
\bibitem{BoutalebJoutei:1979va}
  H.~Boutaleb-Joutei, A.~Chakrabarti and A.~Comtet,
  Phys.\ Rev.\  D {\bf 20} (1979) 1884.
\bibitem{Chakrabarti:1987kz}
  A.~Chakrabarti,
  Fortsch.\ Phys.\  {\bf 35} (1987) 1.
\bibitem{O'Brien:1988rs}
  G.~M.~O'Brien and D.~H.~Tchrakian,
  J.\ Math.\ Phys.\  {\bf 29} (1988) 1212.
\bibitem{Radu:2007az}
  E.~Radu, D.~H.~Tchrakian and Y.~Yang,
  Phys.\ Rev.\  D {\bf 77} (2008) 044017
  [arXiv:0707.1270 [hep-th]].
\bibitem{Belavin:1975fg}
  A.~A.~Belavin, A.~M.~Polyakov, A.~S.~Shvarts and Y.~S.~Tyupkin,
  Phys.\ Lett.\ B {\bf 59} (1975) 85.
\bibitem{Tchrakian:1984gq}
  D.~H.~Tchrakian,
  Phys.\ Lett.\  B {\bf 150} (1985) 360.
\bibitem{Witten:1976ck}
  E.~Witten,
  Phys.\ Rev.\ Lett.\  {\bf 38} (1977) 121.
\bibitem{Chakrabarti:1985qj}
  A.~Chakrabarti, T.~N.~Sherry and D.~H.~Tchrakian,
  Phys.\ Lett.\  B {\bf 162} (1985) 340.
\bibitem{Spruck:1997eb}
  J.~Spruck, D.~H.~Tchrakian and Y.~Yang,
  Commun.\ Math.\ Phys.\  {\bf 188} (1997) 737.
\bibitem{Tchrakian:1990gc}
  D.~H.~Tchrakian and A.~Chakrabarti,
  J.\ Math.\ Phys.\  {\bf 32} (1991) 2532.
\bibitem{O'Se:1987fx}
  D.~O'Se and D.~H.~Tchrakian,
  Lett.\ Math.\ Phys.\  {\bf 13} (1987) 211.
\bibitem{Ma:1990ja}
  Z.~Ma and D.~H.~Tchrakian,
  J.\ Math.\ Phys.\  {\bf 31} (1990) 1506.
\bibitem{Schwarz:1977ix}
  A.~S.~Schwarz,
  Commun.\ Math.\ Phys.\  {\bf 56} (1977) 79.
\bibitem{Romanov:1977rr}
  V.~N.~Romanov, A.~S.~Schwarz and Yu.~S.~Tyupkin,
  Nucl.\ Phys.\  B {\bf 130} (1977) 209.
\bibitem{Schwarz:1981mb}
  A.~S.~Schwarz and Yu.~S.~Tyupkin,
  Nucl.\ Phys.\  B {\bf 187} (1981) 321.
\bibitem{Ma:1986pu}
  Z.~Q.~Ma, G.~M.~O'Brien and D.~H.~Tchrakian,
  Phys.\ Rev.\  D {\bf 33} (1986) 1177.
\bibitem{Ma:1988um}
  Z.~Q.~Ma and D.~H.~Tchrakian,
  Phys.\ Rev.\  D {\bf 38} (1988) 3827.
\bibitem{Radu:2005rf}
  E.~Radu and D.~H.~Tchrakian,
  Phys.\ Rev.\  D {\bf 71} (2005) 125013
  [arXiv:hep-th/0502025].
\bibitem{Julia:1975ff}
  B.~Julia and A.~Zee,
  Phys.\ Rev.\  D {\bf 11} (1975) 2227.
\bibitem{O'Brien:1988xr}
  G.~M.~O'Brien and D.~H.~Tchrakian,
  Mod.\ Phys.\ Lett.\  A {\bf 4} (1989) 1389.
\bibitem{Ibadov:2007qt}
  R.~Ibadov, B.~Kleihaus, J.~Kunz and U.~Neemann,
  Phys.\ Lett.\  B {\bf 659} (2008) 421
  [arXiv:0709.0285 [gr-qc]].
\bibitem{Galtsov:1989ip}
  D.~V.~Galtsov and A.~A.~Ershov,
  Phys.\ Lett.\  A {\bf 138} (1989) 160.
\bibitem{Bizon:1992pi}
  P.~Bizon and O.~T.~Popp,
  Class.\ Quant.\ Grav.\  {\bf 9} (1992) 193.
\bibitem{Kanti:1995vq}
  P.~Kanti, N.~E.~Mavromatos, J.~Rizos, K.~Tamvakis and E.~Winstanley,
  Phys.\ Rev.\  D {\bf 54} (1996) 5049
  [arXiv:hep-th/9511071].
\bibitem{Winstanley:2008ac}
  E.~Winstanley,
  Lect.\ Notes Phys.\  {\bf 769} (2009) 49
  [arXiv:0801.0527 [gr-qc]].
\bibitem{Straumann:1990as}
  N.~Straumann and Z.~H.~Zhou,
  Phys.\ Lett.\  B {\bf 243} (1990) 33.
\bibitem{Bizon:1991hw}
  P.~Bizon,
  Phys.\ Lett.\  B {\bf 259} (1991) 53.
\bibitem{Breitenlohner:2003qj}
  P.~Breitenlohner, D.~Maison and G.~Lavrelashvili,
  Class.\ Quant.\ Grav.\  {\bf 21} (2004) 1667
  [arXiv:gr-qc/0307029].
\bibitem{VanderBij:2001ia}
  J.~J.~Van der Bij and E.~Radu,
  Phys.\ Lett.\  B {\bf 536} (2002) 107
  [arXiv:gr-qc/0107065].
\bibitem{Gubser:2008zu}
  S.~S.~Gubser,
  Phys.\ Rev.\ Lett.\  {\bf 101} (2008) 191601
  [arXiv:0803.3483 [hep-th]].
\bibitem{Breitenlohner:2004fp}
  P.~Breitenlohner, P.~Forgacs and D.~Maison,
  Commun.\ Math.\ Phys.\  {\bf 261} (2006) 569
  [arXiv:gr-qc/0412067].
\bibitem{Brihaye:2006kn}
  Y.~Brihaye, B.~Hartmann, E.~Radu and C.~Stelea,
  Nucl.\ Phys.\  B {\bf 763} (2007) 115
  [arXiv:gr-qc/0607078].
\bibitem{Torii:1995wv}
  T.~Torii, K.~i.~Maeda and T.~Tachizawa,
  Phys.\ Rev.\  D {\bf 52} (1995) 4272
  [arXiv:gr-qc/9506018].
  
\bibitem{Brihaye:2005ft}
  Y.~Brihaye, B.~Hartmann and E.~Radu,
  Phys.\ Rev.\ Lett.\  {\bf 96} (2006) 071101
  [arXiv:hep-th/0508247].
\bibitem{Israel:1967za}
  W.~Israel,
  Commun.\ Math.\ Phys.\  {\bf 8} (1968) 245.
\bibitem{Ibadov:2004rt}
  R.~Ibadov, B.~Kleihaus, J.~Kunz and Y.~Shnir,
  Phys.\ Lett.\  B {\bf 609} (2005) 150
  [arXiv:gr-qc/0410091].
\bibitem{Kleihaus:2005fs}
  B.~Kleihaus, J.~Kunz and U.~Neemann,
  Phys.\ Lett.\  B {\bf 623} (2005) 171
  [arXiv:gr-qc/0507047].
\bibitem{Ibadov:2005rb}
  R.~Ibadov, B.~Kleihaus, J.~Kunz and M.~Wirschins,
  Phys.\ Lett.\  B {\bf 627} (2005) 180
  [arXiv:gr-qc/0507110].
\bibitem{Kleihaus:2007vf}
  B.~Kleihaus, J.~Kunz, F.~Navarro-Lerida and U.~Neemann,
  Gen.\ Rel.\ Grav.\  {\bf 40} (2008) 1279
  [arXiv:0705.1511 [gr-qc]].
  
\bibitem{Radu:2001ij}
  E.~Radu,
  Phys.\ Rev.\  D {\bf 65} (2002) 044005
  [arXiv:gr-qc/0109015].
\bibitem{Radu:2004gu}
  E.~Radu and E.~Winstanley,
  Phys.\ Rev.\  D {\bf 70} (2004) 084023
  [arXiv:hep-th/0407248].
\bibitem{Mann:2006jc}
  R.~B.~Mann, E.~Radu and D.~H.~Tchrakian,
  Phys.\ Rev.\  D {\bf 74} (2006) 064015
  [arXiv:hep-th/0606004].
\bibitem{Kleihaus:2000kg}
  B.~Kleihaus and J.~Kunz,
  Phys.\ Rev.\ Lett.\  {\bf 86} (2001) 3704
  [arXiv:gr-qc/0012081].

\bibitem{Kleihaus:2002ee}
  B.~Kleihaus, J.~Kunz and F.~Navarro-Lerida,
  Phys.\ Rev.\  D {\bf 66} (2002) 104001
  [arXiv:gr-qc/0207042].
\bibitem{VanderBij:2001nm}
  J.~J.~Van der Bij and E.~Radu,
  Int.\ J.\ Mod.\ Phys.\  A {\bf 17} (2002) 1477
  [arXiv:gr-qc/0111046].
 
\bibitem{Brodbeck:1997ek}
  O.~Brodbeck, M.~Heusler, N.~Straumann and M.~S.~Volkov,
  Phys.\ Rev.\ Lett.\  {\bf 79} (1997) 4310
  [arXiv:gr-qc/9707057].
\bibitem{Radu:2002rv}
  E.~Radu,
  Phys.\ Lett.\  B {\bf 548} (2002) 224
  [arXiv:gr-qc/0210074].
\bibitem{Radu:2002hf}
  E.~Radu,
  Phys.\ Rev.\  D {\bf 67} (2003) 084030
  [arXiv:hep-th/0211120].

\bibitem{Brihaye:2005ak}
  Y.~Brihaye and E.~Radu,
  Phys.\ Lett.\  B {\bf 615} (2005) 1
  [arXiv:gr-qc/0502053].
\bibitem{NUT}
E. T. Newman, L.Tamburino and T. Unti, J. Math. Phys. {\bf 4} (1963)
915;
\\
C. W. Misner, J. Math. Phys. {\bf 4} (1963) 924;
\\
C. W. Misner and A. H. Taub, Sov. Phys. JETP {\bf 28} (1969) 122. 
 
\bibitem{Brill}
D.~R.~Brill, Phys.\ Rev. {\bf 133} (1964) B845.
\bibitem{Hertog:2004dr}
  T.~Hertog and K.~Maeda,
  JHEP {\bf 0407} (2004) 051
  [arXiv:hep-th/0404261];
\\
  T.~Hertog and K.~Maeda,
  Phys.\ Rev.\ D {\bf 71} (2005) 024001
  [arXiv:hep-th/0409314];
\\
  M.~Henneaux, C.~Martinez, R.~Troncoso and J.~Zanelli,
  Phys.\ Rev.\ D {\bf 70} (2004) 044034
  [arXiv:hep-th/0404236];
  \\
  E.~Radu and D.~H.~Tchrakian,
  Class.\ Quant.\ Grav.\  {\bf 22} (2005) 879
  [arXiv:hep-th/0410154];
\\
  J.~T.~Liu and W.~A.~Sabra,
  Phys.\ Rev.\  D {\bf 72} (2005) 064021
  [arXiv:hep-th/0405171].
\bibitem{Manvelyan:2008sv}
  R.~Manvelyan, E.~Radu and D.~H.~Tchrakian,
  arXiv:0812.3531 [hep-th].
\bibitem{Brihaye:2007tw}
  Y.~Brihaye, E.~Radu and D.~H.~Tchrakian,
  Phys.\ Rev.\  D {\bf 76} (2007) 105005
  [arXiv:0707.0552 [hep-th]].   
\bibitem{Balasubramanian:1999re}
  V.~Balasubramanian and P.~Kraus,
  Commun.\ Math.\ Phys.\  {\bf 208} (1999) 413
  [arXiv:hep-th/9902121].
\bibitem{yang}
C.~N.~Yang, J. Math. Phys. {\bf 19} (1978) 320.
\bibitem{Tchrakian:2008zz}
  T.~Tchrakian,
  Phys.\ Atom.\ Nucl.\  {\bf 71} (2008) 1116.
\bibitem{Gibbons:2006wd}
  G.~W.~Gibbons and P.~K.~Townsend,
  Class.\ Quant.\ Grav.\  {\bf 23} (2006) 4873
  [arXiv:hep-th/0604024];
\\
  P.~Diaz and A.~Segui,
  Phys.\ Rev.\  D {\bf 76} (2007) 064033
  [arXiv:0704.0366 [gr-qc]];
\\
  A.~Belhaj, P.~Diaz and A.~Segui,
  arXiv:0906.0489 [hep-th];
\\
  S.~H.~Mazharimousavi and M.~Halilsoy,
  Phys.\ Lett.\  B {\bf 659} (2008) 471
  [arXiv:0801.1554 [gr-qc]].
\bibitem{Kleihaus:1998gy}
  B.~Kleihaus, D.~O'Keeffe and D.~H.~Tchrakian,
  Nucl.\ Phys.\  B {\bf 536} (1998) 381
  [arXiv:hep-th/9806088].
\bibitem{Hartmann:2008hq}
  B.~Hartmann,
  arXiv:0811.4076 [gr-qc].
\bibitem{Volkov:2006xt}
  M.~S.~Volkov,
{\it Gravitating non-Abelian solitons and hairy black holes in higher
  dimensions,}
  in {\it "Berlin 2006, Marcel Grossmann Meeting on General Relativity"}, 1379-1396
  [arXiv:hep-th/0612219].
\bibitem{Brihaye:2004kh}
  Y.~Brihaye and E.~Radu,
  Phys.\ Lett.\  B {\bf 605} (2005) 190
  [arXiv:hep-th/0409065].
\bibitem{Hartmann:2004tx}
  B.~Hartmann,
  Phys.\ Lett.\  B {\bf 602} (2004) 231
  [arXiv:hep-th/0409006];
  \\
  Y.~Brihaye, B.~Hartmann and E.~Radu,
  Phys.\ Rev.\  D {\bf 72} (2005) 104008
  [arXiv:hep-th/0508028].
\bibitem{Brihaye:2005pz}
  Y.~Brihaye, B.~Hartmann and E.~Radu,
  Phys.\ Rev.\  D {\bf 71} (2005) 085002
  [arXiv:hep-th/0502131].
\bibitem{Brihaye:2007jua}
  Y.~Brihaye and E.~Radu,
  Phys.\ Lett.\  B {\bf 658} (2008) 164
  [arXiv:0706.4378 [hep-th]].
\bibitem{Pope:1978kj}
  C.~N.~Pope and A.~L.~Yuille,
  Phys.\ Lett.\  B {\bf 78} (1978) 424.
\bibitem{Bianchi:1996zj}
  M.~Bianchi, F.~Fucito, G.~Rossi and M.~Martellini,
  Nucl.\ Phys.\  B {\bf 473} (1996) 367
  [arXiv:hep-th/9601162].
\bibitem{Etesi:2002cc}
  G.~Etesi and T.~Hausel,
  Commun.\ Math.\ Phys.\  {\bf 235} (2003) 275
  [arXiv:hep-th/0207196].
\bibitem{Cherkis:2009jm}
  S.~A.~Cherkis,
  arXiv:0902.4724 [hep-th].
\bibitem{Brihaye:2006nk}
  Y.~Brihaye and E.~Radu,
  Phys.\ Lett.\  B {\bf 636} (2006) 212
  [arXiv:gr-qc/0602069].
\bibitem{Brihaye:2006bk}
  Y.~Brihaye and E.~Radu,
  Europhys.\ Lett.\  {\bf 75} (2006) 730
  [arXiv:hep-th/0605111].
  
\bibitem{Tekin:2002mt}
  B.~Tekin,
  Phys.\ Rev.\  D {\bf 65} (2002) 084035
  [arXiv:hep-th/0201050].
\bibitem{Sarioglu:2009du}
  O.~Sarioglu and B.~Tekin,
  Phys.\ Rev.\  D {\bf 79} (2009) 104024
  [arXiv:0903.3803 [hep-th]].
\bibitem{Maldacena:2004rf}
  J.~M.~Maldacena and L.~Maoz,
  JHEP {\bf 0402} (2004) 053
  [arXiv:hep-th/0401024].
\bibitem{Kihara:2007di}
  H.~Kihara and M.~Nitta,
  Phys.\ Rev.\  D {\bf 77} (2008) 047702
  [arXiv:hep-th/0703166].

\bibitem{Kihara:2007vz}
  H.~Kihara and M.~Nitta,
  Phys.\ Rev.\  D {\bf 76} (2007) 085001
  [arXiv:0704.0505 [hep-th]].
 \bibitem{Tseytlin}
A. A. Tseytlin, {\it Born--Infeld action, suersymmetry and string theory},
in {\it "Yuri Golfand memorial volume"}, ed. M. Shifman, World Scientific, 2000.
\bibitem{BRS}
E. Bergshoeff, M. de Roo and A. Sevrin,
{\bf 49} (2001) 433-440; {\it ibid.} (2001) 50-55.
\bibitem{CNT}
M. Cederwall, B. Nilsson and D. Tsimpis, JHEP 0106 (2001) 034.   
\bibitem{Gauntlett:1992nn} 
  J.~P.~Gauntlett, J.~A.~Harvey and J.~T.~Liu,
  Nucl.\ Phys.\  B {\bf 409} (1993) 363
  [arXiv:hep-th/9211056].
\bibitem{Bergshoeff:2006bs}
  E.~A.~Bergshoeff, G.~W.~Gibbons and P.~K.~Townsend,
  Phys.\ Rev.\ Lett.\  {\bf 97} (2006) 231601
  [arXiv:hep-th/0607193].
\bibitem{Gibbons:1993xt}
  G.~W.~Gibbons, D.~Kastor, L.~A.~J.~London, P.~K.~Townsend and J.~H.~Traschen,
  Nucl.\ Phys.\  B {\bf 416} (1994) 850
  [arXiv:hep-th/9310118].      
\bibitem{Minasian:2001ib}
  R.~Minasian, S.~L.~Shatashvili and P.~Vanhove,
  Nucl.\ Phys.\  B {\bf 613} (2001) 87
  [arXiv:hep-th/0106096].
\bibitem{Polchinski:2005bg}
  J.~Polchinski,
  JHEP {\bf 0609} (2006) 082
  [arXiv:hep-th/0510033].
\bibitem{Emparan:2009cs}
  R.~Emparan, T.~Harmark, V.~Niarchos and N.~A.~Obers,
  Phys.\ Rev.\ Lett.\  {\bf 102} (2009) 191301
  [arXiv:0902.0427 [hep-th]].
 
  
\end{thebibliography}
\end{document}